\documentclass[prd,nofootinbib,floatfix]{revtex4}

\usepackage[british]{babel}
\usepackage{amsbsy}
\usepackage{amssymb}
\usepackage[sumlimits,tbtags]{amsmath}
\usepackage{graphicx}
\usepackage{subcaption}
\usepackage{xcolor}
\usepackage[active]{srcltx}
\usepackage{pdfsync}
\usepackage{xfrac}
\usepackage{svg}

\usepackage{siunitx}
\DeclareSIUnit{\dx}{\ensuremath{\Delta \mathit{x}}}

\usepackage{mathtools}
\usepackage{ulem}

\usepackage{lipsum}
\usepackage{comment}

\usepackage[hyperfootnotes = false]{hyperref}
\usepackage{cleveref}
\hypersetup{
 colorlinks,
 linkcolor=blue,
 citecolor=red,
 urlcolor=blue, 
 }
 
\usepackage{color}

\newcommand{\change}[1]{#1}

\newcommand{\la}{{\langle}}
\newcommand{\ra}{{\rangle}}

\newcommand{\veps}{{\varepsilon}}

\def\O{{\mathcal O}}

\def\tn{{\tilde{n}}}
\def\teps{{\tilde{\varepsilon}}}
\def\tu{{\tilde{u}}}
\def\tsig{{\tilde{\sigma}}}
\def\ttheta{{\tilde{\theta}}}
\def\tvort{{\tilde{\omega}}}
\def\ta{{\tilde{a}}}
\def\tT{{\tilde{T}}}

\def\tmu{{\tilde{\mu}}}

\def\tperp{{\tilde{\perp}}}

\begin{document}

\title{Covariant approach to relativistic large-eddy simulations: Lagrangian filtering}

\author{T. Celora$^1$, M. J. Hatton$^2$, I. Hawke$^2$, N. Andersson$^2$}

\affiliation{
$^1$ Institute of Space Sciences (ICE, CSIC), Campus UAB, Carrer de Magrans, 08193 Barcelona, Spain\\
$^2$ Mathematical Sciences and STAG Research Centre, University of Southampton,
Southampton SO17 1BJ, United Kingdom}

\begin{abstract}
We present a proof-of-principle implementation of the first fully covariant filtering scheme applied to relativistic fluid turbulence.
The filtering is performed with respect to special observers, identified dynamically as moving with the `bulk of the flow'. This means that filtering does not depend on foliations of spacetime but rather on the intrinsic fibration traced out by the observers.
The covariance of the approach means that the results may be lifted into an arbitrary, curved spacetime.
This practical step follows  theoretical work showing that the residuals introduced by filtering a fine-scale ideal fluid can be represented by a non-ideal fluid prescription at the coarse scale.
We interpret such non-ideal terms using a simple first-order gradient model, which allows us to extract effective turbulent viscosities and conductivity. 
A statistical regression on these terms shows that the majority of their variation may be explained by the thermodynamic properties of the filtered fluid and invariants of its flow, such as the shear and vorticity. 
This serves as a validation of the method and enables us to fit a functional, power-law form for the non-ideal coefficients---an approach that may be used practically to give a sub-grid closure model in large-eddy simulations.

\end{abstract}

\maketitle

\section{Introduction}

Since the first binary neutron star merger was detected nearly seven years ago \cite{LIGOScientific_2017_gw170817,LIGOScientific_2017_gw170817multi}, much has been learnt about these extreme objects from the joint observation of gravitational waves and electromagnetic signals. Matching of the observed gravitational waveform to templates created by numerical relativity simulations has placed constraints on the radius, mass, and tidal deformability of neutron stars \cite{raithel2019constraints}. This informs us about the supranuclear equation of state (EoS). Meanwhile, the electromagnetic counterpart signals provide insight into how r-process nucleosynthesis yields many of the heavy elements in our universe \cite{metzger2020kilonovae}. 

Already, people are looking ahead to the next generation of ground-based gravitational wave detectors such as LIGO-voyager~\cite{berti_snowmass2021_2022}, NEMO~\cite{ackley_neutron_2020}, Cosmic Explorer~\cite{reitze_cosmic_2019} and the Einstein Telescope~\cite{punturo_einstein_2010,Maggiore_ET_sciencecase}. These instruments will bring  an order of magnitude improvement in gravitational wave sensitivity, leading to a significant increase in the merger detection rate. Improved sensitivity at higher (kHz) frequencies  should also allow detection of the post-merger signal.  It is in this phase that the subtleties of modelling the remnant's behaviour become important for discerning the precise physics involved. With more precise data comes the need for more accurate numerical simulations to draw meaningful inference from the observations. In particular, the post-merger gravitational-wave emission is sensitive to the small-scale dynamics of nuclear reactions \cite{alford_viscous_2018,hammond_detecting_2022,most_emergence_2022,chabanov2023impact}, electromagnetism \cite{aguilera-miret_role_2023,Most23_dynamo,kiuchi2024_dynamo} and turbulence \cite{radice_calibrated,palenzuela_turbulent_2022}. 

The disparate array of dynamical time and length scales present in binary neutron star mergers presents a major challenge for any numerical code looking to directly resolve them all. This is, in fact,  an impossibility for now and the foreseeable future. To illustrate this quantitatively, consider that the highest simulation resolutions achieved so far have grids with side-length $\mathcal{O}( \qty{10}{\metre})$~\citep{kiuchi_efficient_2015} whilst the molecular dissipation scale in the neutron star fluid is $\mathcal{O}(\qty{1}{\centi\metre})$. Alternatively, the Reynolds number of the flows in neutron stars may range as widely as $\numrange{e6}{e15}$ during a merger event \cite{DavidIanReview}. Nuclear reactions, turbulence and neutrino processes all lead to variations in fluid properties below the resolved length-scale. This violates the fundamental assumption of fluid dynamics: uniformity within a fluid element. Whether we view this discrepancy as stochastic fluctuations or as genuine microphysics may affect our motivations and approach to the problem, but in any case it needs attention.

Turbulence, seeded by Kelvin-Helmholtz instabilities that grow in the shearing layer as  the two stars come into contact, causes a cascade of kinetic energy from the largest length scales down to the smallest, at which point viscous dissipation converts it to heat. 
Unfortunately, the non-linear coupling of scales in turbulent flows means that uncontrolled small-scale effects may spoil the large-scale behaviour. This issue is perhaps epitomized by the unbound growth of maximum magnetic field strength seen in magnetohydrodynamic (MHD) simulations as resolution is increased. As a solution to this problem, large eddy schemes (LES) which include sub-grid closures have been introduced as extensions to the usual general-relativistic MHD  prescription used in simulations. These general relativistic LES (GRLES) models were able to solve the issue of unbound magnetic field amplification and led to saturation instead. Following this, LES schemes are now seeing more widespread usage~\citep{radice2017general, radice_calibrated, vigano_general_2020, carrasco_gradient_2020, palenzuela_large_2022, wang_constant-coefficient_2022, izquierdo_large_2024, aguilera-miret_turbulent_2020,miravet2022_SGS_MRI,miravet2024_SGS_KHI}. For an extensive discussion of the current status of this kind of modelling, see~\cite{DavidIanReview}.

Stripped to the bones, the LES strategy consists of two key steps. First, one introduces a filtering operation with an associated filter kernel and filter length-scale. 
This is done in order to separate a field (or quantity) into a `resolved' part that varies on large scales (that is, scales larger than the filter length-scale) and small-scale fluctuations. 
Second, the filtering operation is applied to the equations of motion. 
The net result is that, whenever the equations of motion contain non-linearities, filtering introduces residual terms that capture the impact of the fluctuation dynamics on the resolved scales. 
By modelling these residuals---e.g. calibrating them on high-resolution simulations of turbulence---one can try to inform a large-scale simulation with relatively low resolution about dynamics happening at smaller scales. 
These steps form the basis of the LES strategy,  in the Newtonian context as well as in relativity.

A significant drawback of all relativistic LES implementations to date is that none  of them are covariant, as highlighted in~\cite{eyink2018cascades,celora_covariant_2021}. This is because the filtering procedure, which takes one from the `micro'-level (where everything is assumed to be known and calculable) to the `meso'-level (where simulations are typically carried out) is performed at the level of the (foliation-based, or 3+1) equations of motion and not on, for example, the stress-energy tensor itself. This makes the procedure frame dependent and so must be performed with respect to a (perhaps implicitly) chosen observer. If we instead choose to set the filtering observer using the fluid's physical state, we can define a unique observer that will give us the same filtering result independently of the frame of reference or coordinates used.  
These issues were discussed in~\cite{duez_comparison_2020} and subsequently addressed in~\cite{celora_covariant_2021}, where an alternative, covariant framework using a fibration was put forward. This, together with the higher-level discussion we provide in \cite{FilteringStrategy}, serves as the theoretical precursor to the present more practical work.

In this paper we implement the fibration-based, fully covariant, LES strategy outlined in~\cite{celora_covariant_2021}. We lay out the practical logic in \cref{sec:logic} and apply the scheme to real numerical data in \cref{sec:KHcase}.
We then build a closure model in \cref{sec:DiscriminatingModels} and consider the impact on the interpretation of the micro-physical parameters in \cref{sec:eos_check}. The main purpose of this paper is to illustrate the viability of the approach and explore the possible pitfalls of alternative strategies.

\section{Lagrangian filtering: logic of the scheme}\label{sec:logic}

We now discuss a practical implementation of the framework presented in \cite{celora_covariant_2021}, focusing on the logic of the scheme and deferring to \cref{sec:KHcase} a detailed proof of principle analysis. 
We will here apply the scheme to special relativistic box simulations of hydrodynamic turbulence, even though we anticipate extensions to magnetized systems, as discussed in \cite{FilteringStrategy}. 
The covariance of the Lagrangian filtering and the fact that this leaves the metric unaffected, allow us to apply the scheme to special relativistic numerical data and lift the results of our analysis to any spacetime. 
In principle, this allows us to perform explicit large-eddy scheme simulations in curved spacetimes---as required for binary neutron star merger simulations---using a sub-grid model fitted within our framework. 
Of course, re-calibration to the specific dynamics of other systems beyond those of our box simulations here may be required for this to be accurate.

The first cornerstone of the scheme is the construction of the observer $U^a$, which will be used for filtering and post-processing the numerical data.
In essence, the idea is to work around the fact that any filtering operation breaks covariance by linking the operation to a physically meaningful observer---not to a gauge-dependent foliation observer. 
Then, introducing Fermi-coordinates (see, e.g. \cite{GravitationMTW}) associated with such an observer, one can show that the metric is unaffected by the filtering operation. 
These two features combined mean that the framework retains compatibility with the tenets of general relativity, while adopting the main ideas of the more traditional LES strategy. We do not dwell on these issues here as they are extensively discussed in \cite{celora_covariant_2021} and \cite{FilteringStrategy}.

In practice, as for turbulent flows we expect to have changes in the flow properties over small scales, we can introduce the filtering observer as the one that moves with the `bulk' of the fluid. That is to say the observer is aligned with the (baryon) number current of the flow, at the coarse level.
This means that, if we consider a `box' in spacetime that is adapted to such an observer---that is, the observer identifies the time-like direction of the box---there should be no (or at least minimized) baryon flux through opposite sides of the box. 
To `extract' the filtering observer from the numerical data, then, we start by constructing such a box around a grid-point using the fine-scale fluid velocity at the point, $u^a$, as initial guess for the observer: $U^a = u^a$. 
Starting from an orthonormal basis for the foliation of the spacetime $\{e_{(t)}^a,e_{(x)}^a,e_{(y)}^a,e_{(z)}^a\} =  \{(1,0,0,0),(0,1,0,0),(0,0,1,0),(0,0,0,1)\}$, we project first spatial leg of the tetrad as 
\begin{equation}
    E^a_{(1)} = e^a_{(x)} + U^a U_b e^b_{(x)} \;, \quad E^a_{(1)} E_a^{(1)} =1 \;.
\end{equation}
The remaining two spatial legs can be projected in a similar fashion, namely
\begin{equation}
    E^a_{(2)} = e^a_{(y)} + U^a U_b e^b_{(y)} - E^a_{(1)}E_b^{(1)}e^b_{(y)} \;, \quad E^a_{(2)} E_a^{(2)} =1 \;,
\end{equation}
and similarly for $E^a_{(3)}$.
Given the tetrad legs, we then consider a space-time volume $\mathcal{V}_{L}$ aligned with the tetrad $\{U^a\,,E_{(1)}^a\,,E_{(2)}^a\,,E_{(3)}^a\}$ with representative lengthscale $L$ about our current point and compute three residual terms
\begin{equation}
    r_{(I)} = \int_{\mathcal{V}_{L}} E^a_{(I)}n_a\,d\mathcal{V}_{L} \;,\quad I=1,2,3
\end{equation}
where $n^a = nu^a$ is the micro-scale baryon current. 
These residuals measure the average particle drift in the direction $I$ over the space-time volume $\mathcal V_{L}$.  
We can then feed these residuals into a root-finding algorithm and reconstruct (from the roots) the observer $U^a$ that minimizes the particle drift on average\footnote{We note that we have also implemented and tested a minimization algorithm based on a similar logic, but found this was slower and more sensitive to the initial guess.}.

Having discussed the construction of the filtering observer, let us  turn to the meso-model and filtering. 
Given discrete data from a box simulation, the first step is to set up a meso-grid that is aligned with the `micro one' from the simulations\footnote{We note that it is not strictly necessary to have the two grids aligned, but this clearly is a sensible choice, particularly for comparison at each grid point.}. 
We do so by introducing a coarse-graining factor (CG) in order to be able to reduce explicitly (or not) the resolution of the meso-grid with respect to the micro-grid. 
Once the grid is set up, we compute the filtering observer at each point on the meso-grid using the procedure discussed above. 
Then, we perform a Lagrangian spatial filtering of all the relevant variable of the micro-model, where we stress that we filter in the spatial directions relative to the observer. 
Here we do this using a sharp box filter kernel, so that for a given quantity $X$ we compute its filtered counterpart as
\begin{equation}
    \la X \ra = \int_{V_{L}} X\, dV_{L} \;,
\end{equation}
where $V_L$ is a spatial volume adapted to $\{E_{(1)}^a\,,E_{(2)}^a\,,E_{(3)}^a\}$ with side $L$ and the integral is performed using a Gauss-Legendre quadrature scheme. 
We note that, as the spatial directions relative to the filtering observer will in general be tilted with respect to the foliation, the procedure requires introducing abstract coordinates adapted to the tilted box at each point.
This operation populates the meso-grid with the computed values for the Lagrangian-filtered meso-model variables. 
For the simplest case of an ideal, single fluid, which we will be focusing on in the following sections, this involves computing the baryon current $\la n^a\ra$, stress-energy tensor $\la T^{ab} \ra$ and thermodynamic pressure $\la p \ra$.

As extensively discussed in \cite{celora_covariant_2021}, even if the micro-scale model is that of an ideal fluid with stress-energy tensor given by
\begin{equation}
    T^{ab} = \varepsilon u^a u^b + p (g^{ab} + u^a u^b)\;,
\end{equation}
where $\varepsilon,\,p$ are the micro-scale energy density and pressure, filtering will introduce additional `non-diagonal' terms into the stress-energy tensor akin to those that enter dissipative fluid models. 
In practice, we introduce a `Favre' observer $\tilde u^a$ at the meso scale as the one associated with the filtered baryon flux $\la n^a\ra \equiv \tilde n \tilde u^a$ and write the filtered stress-energy tensor as 
\begin{equation}\label{eq:FilteredTab_decomposition}
    \la T^{ab}\ra = \left( \tilde \veps + \la p \ra\right) \tilde u^a \tilde u^b +  \la p \ra g^{ab} + 2\tilde q^{(a} \tilde u^{b)} + \tilde s^{ab} \;.
\end{equation}
The tilde notation, $\tilde{\cdot}$, indicates that the variable is a meso-model quantity that may be conceptually related to a micro-model, physically meaningful, variable---we refer to \cite{FilteringStrategy} where this notational choice is explained in detail.
The energy density as measured by the Favre observer is denoted as $\tilde \veps$, while the additional residual terms $\tilde q^a $ and $\tilde s^{ab}$ measure effective momentum flux and stresses introduced by the filtering. 
In essence, these quantities are intended to capture energy and momentum transfer to/from the smaller scales that have been filtered out explicitly. 
Specifying a closure scheme in this picture amounts to introducing a suitable representation of these residuals in terms of resolved/filtered quantities. 
Given the evident analogy with non-dissipative fluids and based on the classical work by Smagorinsky \cite{Smagorinksy} (see also \cite{radice2017general,radice_calibrated}), in \cite{celora_covariant_2021} we proposed a closure scheme that builds precisely on this formal connection. 
Here we choose to model the residuals as 
\begin{subequations}\label{eq:EckartClosure}
\begin{align}
    \tilde s^{ab} &=\frac{1}{3}\tilde \Pi \tilde\perp^{ab} + \tilde\pi^{ab}\;, \\
    \tilde \Pi &=- \zeta \tilde \theta \;, \\
    \tilde \pi^{ab} &= - \eta \tilde \sigma^{ab} \;, \\
    \tilde q^a &= - \kappa \tilde T \tilde\perp^{ab}\left(\frac{1}{\tilde T}\nabla_b \tilde T + \tilde a_b \right) \;,
\end{align}
\end{subequations}
where $\tT$ is the meso temperature,
$\tilde\perp^{ab} = g^{ab} + \tilde u^a \tilde u^b$ is the orthogonal projector relative to the Favre observer, while $\tilde\theta, \tilde\sigma^{ab}$ and $\tilde a^a$ are, respectively, the expansion rate scalar, the shear rate tensor and the acceleration relative to the Favre observer: 
\begin{subequations}\label{subeq:Favre_gradients}
\begin{align}
    \ta_a &= \tu^b\nabla_b\tu_a \;,\\
    \ttheta &= \tperp^{ab}\nabla_a \tu_b \;,\\
    \tsig_{ab} &=\frac{1}{2}\left(\tperp^c_{a} \tperp^d_{b} + \tperp^c_{b} \tperp^d_{a} \right) \nabla_c \tu_d -\frac{1}{3}\ttheta \tperp_{ab}  \;.
\end{align}
\end{subequations}
As extensively discussed in \cite{celora_covariant_2021,DavidIanReview}, filtering may also impact on the thermodynamics. As the equation of state can be thought of as a non-linear closure, there is no guarantee that its functional form will be preserved by filtering. 
This implies that there is some freedom in specifying the EoS and related thermodynamic quantities at the filtered level---see \cite{FilteringStrategy} for a discussion of the repercussions of this.
We do not wish to expand further on this issue here; it will be discussed later in~\cref{sec:eos_check}.
Nonetheless, we note at this point that this may impact on the closure scheme in \cref{eq:EckartClosure} as it requires the introduction and choice of a filtered EoS from which we can compute $\tT$ given filtered data.  

At this point, we have covered the salient points of the scheme, so we can turn our attention to a proof-of-concept implementation of the strategy. 
In fact, the aim of the next sections is to discuss how the effective dissipative coefficients can be extracted from box simulations and how we can use the results to calibrate a meso-model. 
Once this is done, the Lagrangian sub-grid model is complete.  
Before we move on, however, we need to comment on the choice of the closure scheme that will be explored in the following sections.
At first glance, and with the closure scheme above, the final model resembles the Eckart model for a dissipative fluid, which is well-known to be acausal and suffer from instabilities \cite{NilsGregRev, RezzollaZanotti}.
As far as we are aware, however, all discussions of the stability and causality properties of the Eckart model assume constant transport coefficients. It is not clear what would happen to the `classical' results if we relax this assumption. 
In fact, as we will see below, it appears that modelling the effective dissipative coefficients only in terms of thermodynamic quantities like temperature and density does not suffice. 
This casts doubts on the naïve extension of the classical instability results to the problems we are interested in.
Moreover, if the unstable wavenumbers are not resolved in the coarse-grained simulation, then instability would not be triggered at all. We note that a similar perspective has been recently discussed in the context of reaction-sourced bulk-viscosity in mergers \cite{BVinSim}.
On top of all these considerations, it is also important to note that the various models designed to overcome these issues are all built by extending or generalizing the first-order models of the Landau-Eckart family.
Given the novelty of the strategy that is being explored and implemented in this work, and as a first stab at the problem, we will focus on extracting and modelling the effective transport coefficients as in \cref{eq:EckartClosure}, which we may also consider as leading order results.

\section{The Kelvin-Helmoltz case}\label{sec:KHcase}

The Kelvin-Helmholtz instability (KHI) is a shearing instability that results when two (or more) fluid regions flow in opposite directions past each other, usually differing in density. For further details see the recent discussion in~\cite{celora2024MagnetoShear}. A wide range of fluid behaviour can be observed depending on the precise initial data. In arguably the most interesting cases---at least for us---there is an initial linear growth phase of the instability at the interface, followed by a non-linear phase where the creation of vortices and a complex network of shocks typically precedes the onset of smaller-scale turbulence.
Neutron star matter in mergers is likely to be Kelvin-Helmholtz unstable as the two objects collide and shearing flows develop. The KHI is known to play an important role in post-merger dynamics where it moderates the cascade of energy between macroscopic and microscopic scales through the action of shear viscosity in the fluid. This is important in the spin-down of the remnant where the rotational energy of the fluid is converted to small-scale turbulence and then to magnetic energy through the dynamo effect. 

\subsection{Initial data and box simulations}

We now turn our attention to applying the practical filtering scheme laid out in~\cite{celora_covariant_2021} to real simulation data. These data are produced using the METHOD codebase; a (3+1D) finite-difference codebase for simulating magneto-hydrodynamics in special relativistic spacetimes. It contains a range of explicit and implicit-explicit (IMEX) time-integrators, spatial reconstruction and flux schemes. The METHOD codebase has been used previously to simulate resistive magnetohydrodynamics~\cite{wright_resistive_2018,wright_non-ideal_2020,wright_resistive_2020} and dissipative, relativistic fluids~\cite{Hatton2024DEIFY}. For our purposes here, we perform high-resolution simulations of Kelvin-Helmholtz instabilities using a relativistic, ideal fluid description prescribed by the Euler equations. 
The data are defined within a 2D `box' domain where $x \in  [0.0,1.0]; y \in [0.0,1.0]$. The domain is filled with two fluid regions of differing densities that flow past each other with velocities directed in the positive and negative $y$-directions. The inner region of lighter fluid exists roughly in $x \in  [0.25,0.75]$, whilst the outer region of heavier fluid exists elsewhere. 
The primitive variables are
the density and the $y$-component of the velocity:%
\begin{equation}
    \begin{pmatrix}
        \rho \\
        v_y  
    \end{pmatrix} = 
    \begin{dcases}
        \begin{pmatrix}
            0.1 \\
            0.5    
        \end{pmatrix} & x_L < x < x_R, \\
        \begin{pmatrix}
            1.0 \\
            -0.5 
        \end{pmatrix} & \text{otherwise}.
    \end{dcases}
\end{equation}
The interfaces between the two regions---located at $x=x_L \simeq 0.25$ and $x=x_R \simeq 0.75$---are perturbed in a random fashion to induce mixing of the two fluid regions and encourage the instability to grow.
We do so by superimposing $10$ different modes that are out of phase and of different amplitudes. 
Specifically, given $N=1,\dots,10$, we independently perturb the $x$-position of both interfaces as 
\begin{equation}
    x \to x + 0.01 R_{1,N} \cos(R_{2,N} + 2 N \pi x)
\end{equation}
where $R_{1,N}$ are ten random numbers between $0$ and $1$, normalized by their sum, and $R_{2,N}$ are ten random numbers between $-\pi$ and $\pi$.
For the simulations we use a `$\Gamma$-law' EoS of the form:
\begin{equation}\label{eq:GammaLawEoS}
    p = (\Gamma -1)n\epsilon = (\Gamma -1)(\veps - n) \;, \qquad \veps = n(1 + \epsilon)
\end{equation}
where $p$ is the pressure, $\epsilon$ and $\veps$ are (respectively) the specific internal energy and the energy density, while $n$ is the baryon number density. 
The initial pressure is uniform, $p = 1.0$, and the adiabatic index is set to $\Gamma = 4/3$ as for an ultra-relativistic ideal fluid \cite{RezzollaZanotti}. We use periodic boundaries in all directions. Whilst the simulations performed here, and the data analysed from them, are spatially two-dimensional, the filtering codebase is written to work in arbitrary spatial dimensions. The standard simulation resolution used below is $N_x = N_y = 800$.

\begin{figure}[!htb]
    \includegraphics[scale=0.6]{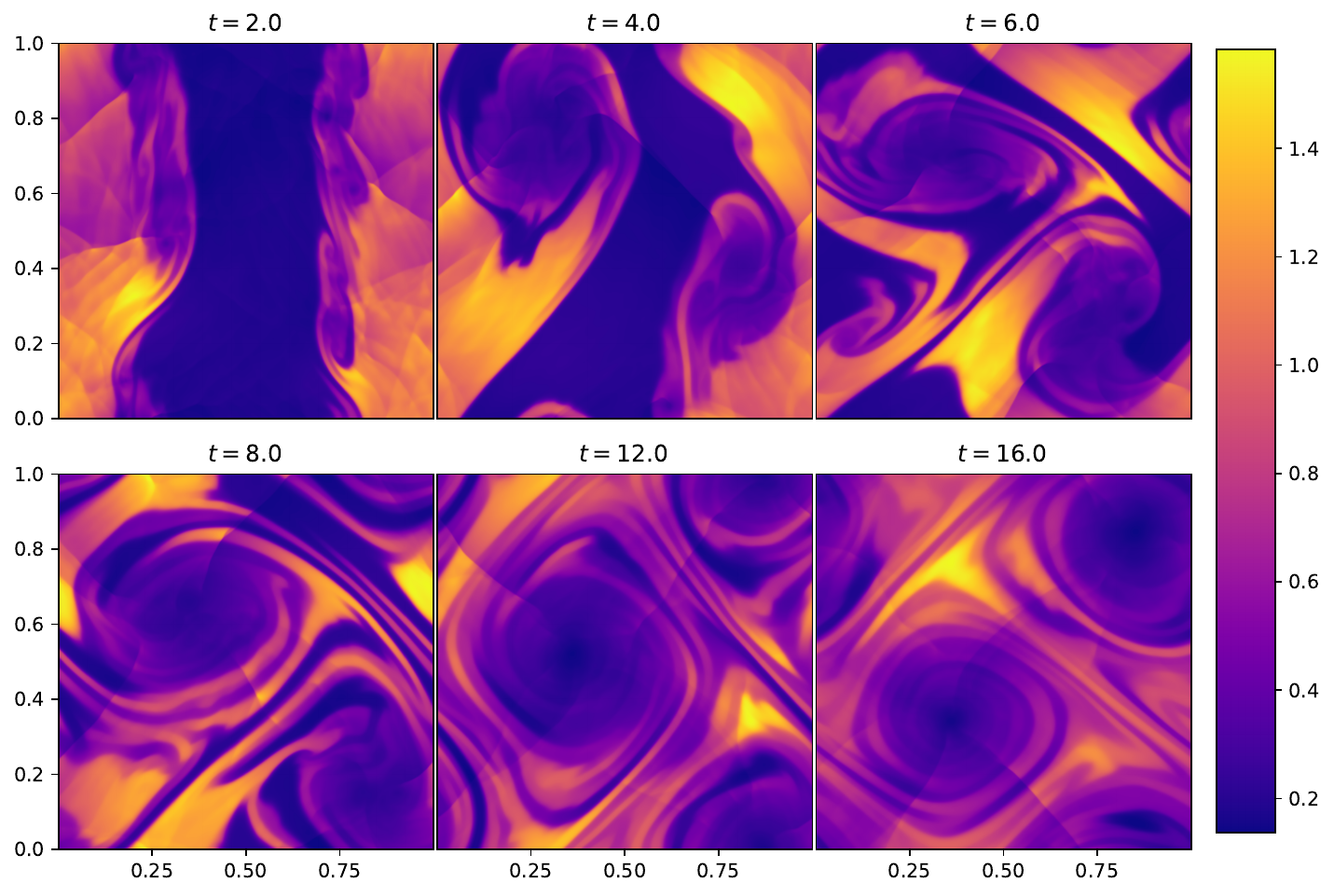}
    \caption{The development of a Kelvin-Helmholtz instability for an ideal fluid with negligible viscosity. The number density is plotted and the simulation ran until a code time of $t=16.0$. Breakdown of the interface, seeded by a random initial perturbation, occurs quickly and leads to vortex formation and the onset turbulence across the domain.}
    \label{fig:KHI_development_n}
\end{figure}

\begin{figure}[!htb]
    \centering
    \begin{minipage}{.5\textwidth}
        \centering
            \includegraphics[scale=0.7]{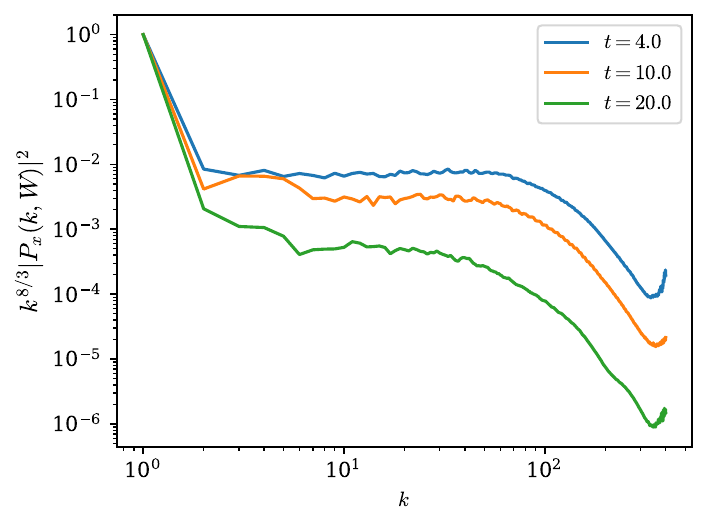}
    \end{minipage}%
    \begin{minipage}{0.5\textwidth}
        \centering
            \includegraphics[scale=0.7]{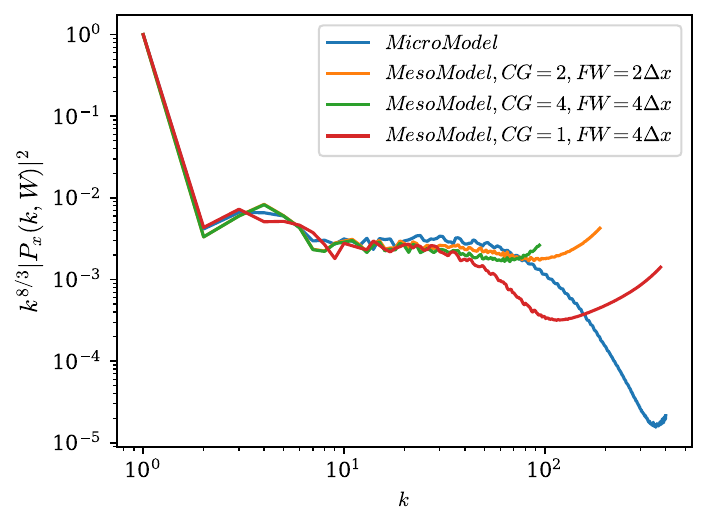}
    \end{minipage}
\caption{The square of the  integrated power spectra of the Lorentz factor for fluid in the Kelvin-Helmholtz simulation seen above in~\cref{fig:KHI_development_n}, adjusted by a prefactor of $k^{8/3}$. Horizontal lines indicate being in the inertial range. On the left, three curves for times $t=4.0$, $t=10.0$ and $t=20.0$ are plotted for the micromodel data (straight from simulation). On the right, four spectra are plotted, all at $t=10.0$: one for the micromodel data again, and three for the mesomodel with varying parameters. CG is the coarse-graining factor of the mesomodel's grid compared to the micromodel's one and FW is the filter width used for the averaging operation. In all cases the inertial range is seen to exist over at least an order of magnitude of wavenumbers.}
    \label{fig:KE_Spectra}
\end{figure} 
When storing METHOD output data for our filtering pipeline, we typically store a series of snapshots (up to $\approx 32$) in rapid succession around a central time-slice. These foliations cover (discretely) a portion of spacetime which we are then able to work within. 
This is required, for example, when calculating the filtering observer at a particular spacetime point, as we need to access data outside the corresponding snapshot.
Similarly, when we perform filtering of the microdata on a particularly slice to obtain the mesodata, we filter over an $n$-dimensional box orthogonal with respect to the filtering observer. This box will project beyond that particular slice and will also be tilted. Hence, we need additional snapshots around those we strictly filter on. The number we need depends on the filter size $L$ which typically falls in the range $\qtyrange{2}{8}{\dx}$.

Once we have obtained these mesodata, we need to calculate the non-ideal terms in our sub-grid closure scheme from them. This will require taking derivatives of the filtered quantities. Spatial derivatives are straightforward to calculate across the meso-grid. However, to calculate temporal derivatives, we again require multiple timeslices of meso-data. Because we are performing covariant filtering, we want to keep time and space on an equal footing. We also want to take derivatives at nearby points in spacetime on fluid with similar global properties. For these reasons, we choose the spatial and temporal capture resolutions to match\footnote{It is not strictly required to have the time-gap in between timeslices exactly equal to the grid-spacing. 
This choice, however, is practical as it means that derivatives in space and time evaluated using the same differencing scheme will have the same level of accuracy.} so that $\Delta t = \Delta x = \Delta y$. This property is maintained when we filter, so that if the resolution of the mesodata is half that of the microdata, this is true in the temporal direction, too. For example, when performing a high-resolution simulation we have $\Delta t = \Delta x = \Delta y = 0.00125$ for the simulation microdata. If we then filter with a coarse-graining factor of $CG=4$, we will choose every fourth timeslice of microdata to work with, such that, $\Delta t_{meso} = \Delta x_{meso} = \Delta y_{meso} = CG \,\Delta t_{micro} = 0.005$ for the mesomodel grid. This demonstrates the need for many initial snapshots, so that even after successive trimmings, we have sufficient data at the end to do statistics on.
Using simple arguments we can estimate how much data we should need, finding that these should cover a total time-interval of about\footnote{Given a central slice, one needs to move in time about $CG \Delta x$ to reach the next useful meso-grid slice, and then up to $L$ beyond that during the observer root-finding step. 
Because we will work with $CG = L/\Delta x$ and evaluate derivatives using a second order centred finite differencing scheme, we need up to three slices and to span a total time-interval of $\approx 4L$.} $4L$.
Given that the maximum value for this is $L=\qty{8}{\dx}$ and that we store snapshots separated by a time-gap equal to $\Delta x$, this gives a maximum total number of snapshots of $\approx 32$ as reported above. 

\Cref{fig:KHI_development_n} shows the evolution of the number density across time for the Kelvin-Helmholtz initial data described above. An asymmetric initial perturbation in the $x$-directed velocity across the interface leads to mixing of the two fluid regions, which differ in density and velocity. This causes large vortices to form which quickly destroy the distinction between the two fluid regions. In time, small-scale structure develops as kinetic energy cascades down the length scales within the simulation.

\Cref{fig:KE_Spectra} summarises results of the integrated power spectra across scales, both for the micromodel (simulation) data in the left panel and for filtered (also sometimes coarse-grained) data in the right panel. For this Fourier analysis, we follow the approach laid out in~\cite{beckwith_second_2011} which is also seen in~\cite{wright_resistive_2018,wright_resistive_2020}.
The results for the micromodel are plotted for three different times so that one may see the features develop. Particularly, a wide inertial range already exists at $t=4.0$, indicated by the flatness of the blue curve.
The inertial range shrinks in time as numerical viscosity plays an increasingly important role in dissipating kinetic energy, which is also shown by the overall lowering of the curves.
The results for the mesomodel are plotted together with one instance of the micromodel data in the right panel. A number of effects shown by the three mesomodel curves  are worth highlighting. 
Firstly, their general shape is preserved with respect to the micromodel, at least at low and medium wavenumbers. This is reassuring as we do not expect a large statistical impact on scales well above that which we are filtering at.
Secondly, the coarse-graining reduces the maximum wavenumber of modes present, which in turn reduces the extent of the spectrum.
Finally, the filtering removes the steep drop-off in kinetic energy seen for the highest wavenumbers in the micromodel data. This is logically what we expect---averaging over small-scale behaviour will remove energy from these scales.
However, we also see a long rising tail in the spectrum (red curve) where this filtering is not combined with matched coarse-graining. In this case, it appears energy has actually been shifted into the smallest scales.
This is an aliasing affect, where the filtering operation maps some power at low frequencies to the highest resolvable frequencies and vice versa. As there is much more power at low frequencies, this leads to higher power at grid size frequencies. This is not problematic for our analysis as our focus is on capturing the behaviour in the inertial range. However, to minimise this effect we will below make the filtering and coarse-graining widths equal.

\subsection{The observers and filtered data}

Having discussed METHOD and the initial data used for the box simulations, let us turn our attention to the observers and filtering. 
The first step is to reconstruct the baryon current and stress-energy tensor as these quantities are typically not directly evolved in numerical simulations. 
Starting from the primitive/auxiliary variables and for the ideal fluid case under consideration, these can be obtained as
\begin{subequations}
\begin{align}
    n^a &= n u^a\;, \quad u^a = \left( W ,\, W v_x ,\, W v_y \right) \;, \\
    T^{ab} & = n h u^a u^b + p \eta^{ab}
\end{align}
\end{subequations}
where $W$ is the Lorentz factor, $v_x,v_y$ the primitive velocities in the $x$ or $y$ direction, $n$ is the baryon density, $h$ the specific enthalpy, $p$ the fluid pressure and $\eta^{ab}$ denotes the Minkowski metric. 
We stress that it is crucial to reconstruct these tensorial quantities from ones that are more conveniently evolved in a hydrodynamic code.
The reason for this is that covariance of our Lagrangian scheme follows from filtering independently each component of a geometrically well-defined object, i.e. a tensor, and use these to reconstruct the tensor at the filtered scale \cite{celora_covariant_2021}. 
As such, for example, we cannot directly apply the filtering operation to the primitive velocities ($v_x$, $v_y$) as these are not the components of a tensor. 

\begin{figure}
    \centering
    \includegraphics[width = 0.95 \textwidth]{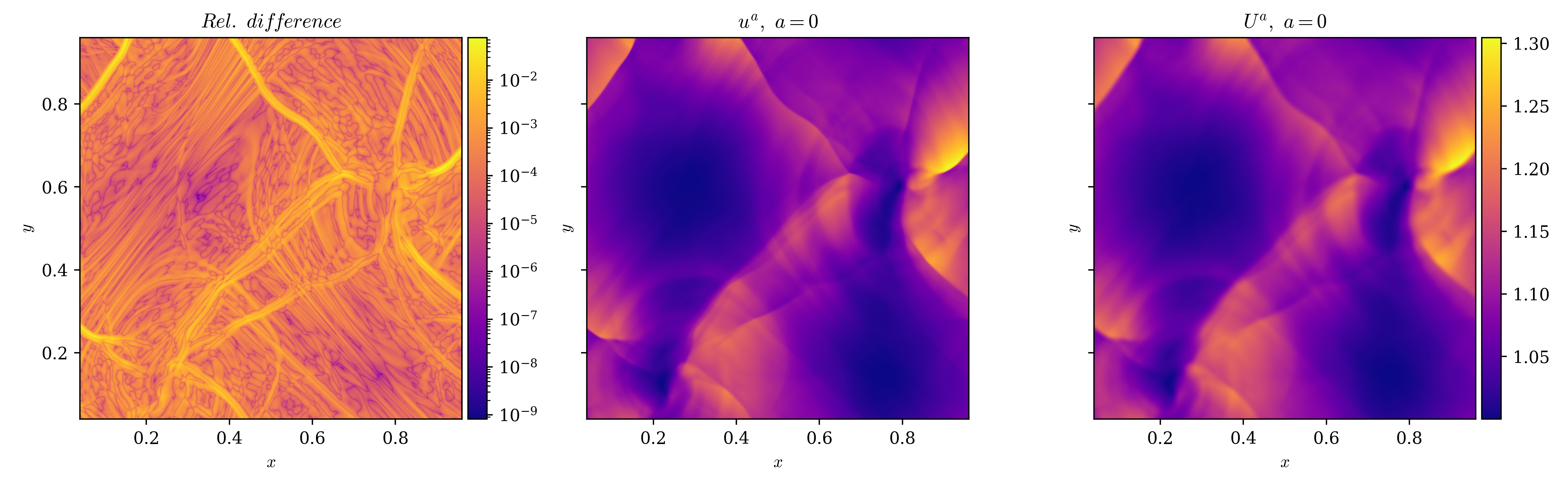}
    \caption{Comparison between the fine-scale velocity $u^a$ (middle panel) and the filtering observers $U^a$ (right panel), computed using a box-length of $\qty{8}{\dx}$ and focusing on the time component. Data from a snapshot taken at around $t=10$. In the left-most panel we show the relative difference between the two velocities, noting that this is typically of the order of a few percent and peaks where $u^a$ presents sharp gradients. }
    \label{fig:ObsVSmicro}
\end{figure}

\begin{figure}
    \centering
    \includegraphics[width = 0.95 \textwidth]{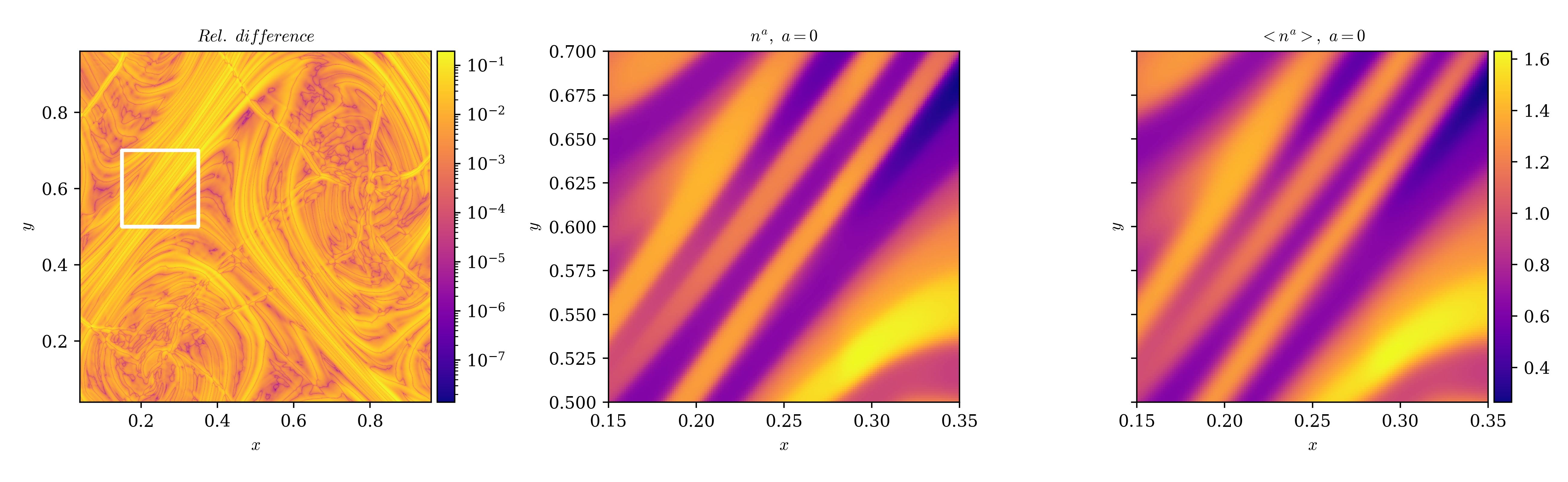}
    \caption{Comparison between the fine-scale $n^a$ and filtered $\la n^a\ra$ baryon current (filter-size $L=\qty{8}{\dx}$), focusing on the time component, at a representative time $t=10$. 
    The left panel shows the relative difference between the two over the full grid. 
    In the middle and right panel we plot the fine-scale and filtered data zooming in the region $x \in (0.15, 0.35)$, $y\in(0.5, 0.7)$ (the box indicated in the left panel). We do so to visually appreciate the effects of filtering: the right panel presents the same `bands' as the middle one but the boundaries between them are smeared out.}\label{fig:FilteringExampleBC}
\end{figure}

Assuming we have computed the baryon current $n^a$ using the available data from the box simulations, we have all the ingredients we need to find the filtering observers following the logic detailed in \cref{sec:logic}. 
We demonstrate this by comparing, in \cref{fig:ObsVSmicro}, the fine-scale velocity $u^a$ against the observers $U^a$ obtained with $L = \qty{8}{\dx}$, focusing on the time component. Data for this figure come from a snapshot around $t=10.0$.
In the middle and right panels we plot data over the entire grid, hence the differences between the two are mainly visible where the fine-scale data shows sharp gradients.
As such, and in order to better compare them, we also plot their relative difference in the left-most panel.
We can then confirm that the relative difference peaks where there are sharp gradients in the fine-scale velocity, and that typical values are of the order of a few percent. 
Let us also note that in constructing similar plots for the $x-y$ components, we tend to observe slightly higher relative differences of the order of $10\%$. 
Nonetheless, we decided to focus here on the Lorentz factor as this gives us a measure for how tilted we may expect the filtering box to be with respect to the foliation\footnote{Since the filtering axes are tilted with respect to the foliation, an interval of $\delta x'$ according to the filtering observer would correspond roughly to a time interval in the foliation of $\delta t = \sqrt{1 - 1/W^2} \delta x'$.}. 
Finally, we tested the robustness of the algorithm used to find the observers against the choice of initial guess. 
To do so, we ran the algorithm with $200$ different initial velocities, obtained by rotating the fine-scale velocity at the point, and looked at the distributions of the observers found. 
We repeated the process at several hundred points randomly selected on the grid. 
In each of these cases, the distributions of the observers found is very well described by a delta-function\footnote{More precisely, we find standard deviations of \numrange{11}{12} orders of magnitude smaller than the mean. Consistently with this, we used a Gaussian kernel density estimator to estimate the observers probability density function, finding this to be clearly limiting to a delta function.} thus showing insensitivity to the initial guess.
This fact, together with the clear physical interpretation of the algorithm implemented, is of crucial importance for the covariance of the scheme.
We now focus our attention on the impact of the Lagrangian filtering. 
As an example, we show in \cref{fig:FilteringExampleBC} the time component of the micro-scale and filtered baryon current. 
In order to appreciate visually the impact of filtering, in the right and middle panel we zoom in on the region  $x \in (0.5, 0.7),\; y\in(0.15, 0.35)$.
The effect of filtering is visible as it smooths out the features present in the fine-scale data. 
As the zoomed in patch present several `bands' of different densities, the effects of filtering are particularly evident in that it smears out the `boundaries' between them.
In addition, the left-most panel of the figure shows the relative difference between fine-scale and filtered data over the full grid. We then observe that typical values range from a few percent up to $10\%$.

\begin{figure}
    \centering
    \includegraphics[width = 1 \textwidth]{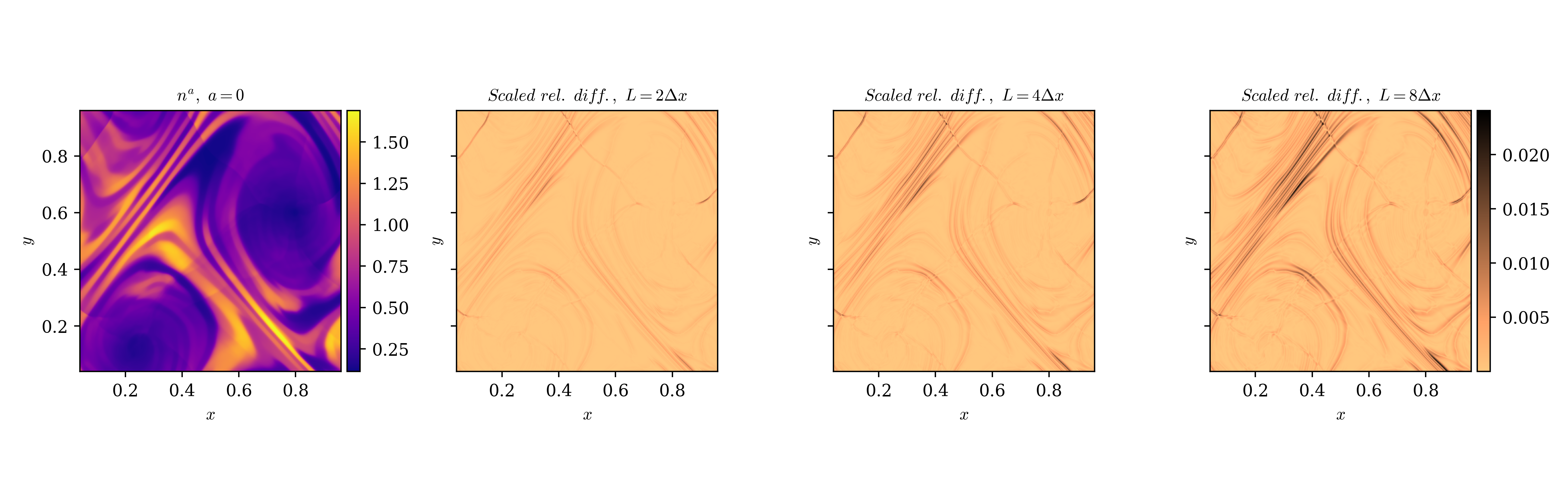}
    \caption{Comparison between the fine-scale $n^a$ and filtered $\la n^a\ra$ baryon current, focusing on the time component, at a representative time $t=10$. 
    The left-most panel shows the fine-scale data, whilst in the panels to its right we plot the relative difference between this and the filtered data---with filter sizes of $\qtylist{2;4;8}{\dx}$. 
    As we observed the maximum values of these to roughly double as we double the filter-size, we plot here the re-scaled relative differences, that is we divide by $\tilde L = L / \Delta x$. 
    This allows us to use a single colour map for the three panels to the right and provides a better comparison among them.
    }
    \label{fig:FilterScalingBC}
\end{figure}

Larger filter-size means interpolating (or combining more generally) data coming from larger regions around a point. 
As such, when comparing filtered meso-scale data to the micro-scale, we expect larger differences for larger filter sizes. 
This would follow automatically had we filtered in the spatial directions identified by the foliation, but such a statement may not be that trivial given the Lagrangian filtering explored in this work. 
In a sense, this provides a useful sanity check of the strategy and the pipeline implemented. 
To verify this, we compared the relative differences between filtered and fine-scale data at various filter widths and observed that the maximum relative difference appears to (roughly) double as we double the filter size. 
In \cref{fig:FilterScalingBC} we show the (time-component of the) fine-scale baryon current $n^a$ and its relative difference for data filtered with filter lengths of $\qtylist{2;4;8}{\dx}$.
Given the observed scaling in the maximum value---this is about $\qtyrange{4}{5}{\percent} $ for $L=\qty{2}{\dx}$, $\qtyrange{8}{10}{\percent}$ for $L=\qty{4}{\dx}$ and up to $\qtyrange{15}{20}{\percent}$ for $L=\qty{8}{\dx}$---we plot the relative difference re-scaled by the filter size $L$.
Even though the impact of filtering is less visible in this figure due to the linear scaling of the colour-map, we confirm the anticipated increase  with the filter size.
We will not try to make a more precise statement at this point given that i) we will get back to this issue in \cref{sec:DiscriminatingModels} ii) the aim here is merely to demonstrate an increasing impact of filtering in a loose sense.  
Moreover, we note that here we have varied the filter size while keeping the observers fixed.
For these tests, we have also constructed the grid for the filtered data as having the \emph{same} number of points as the fine-scale data. 
We stress that we chose to do that in order to focus on the impact of the explicit filtering operation implemented. 

\begin{figure}
    \centering
    \includegraphics[width = 1.0 \textwidth]{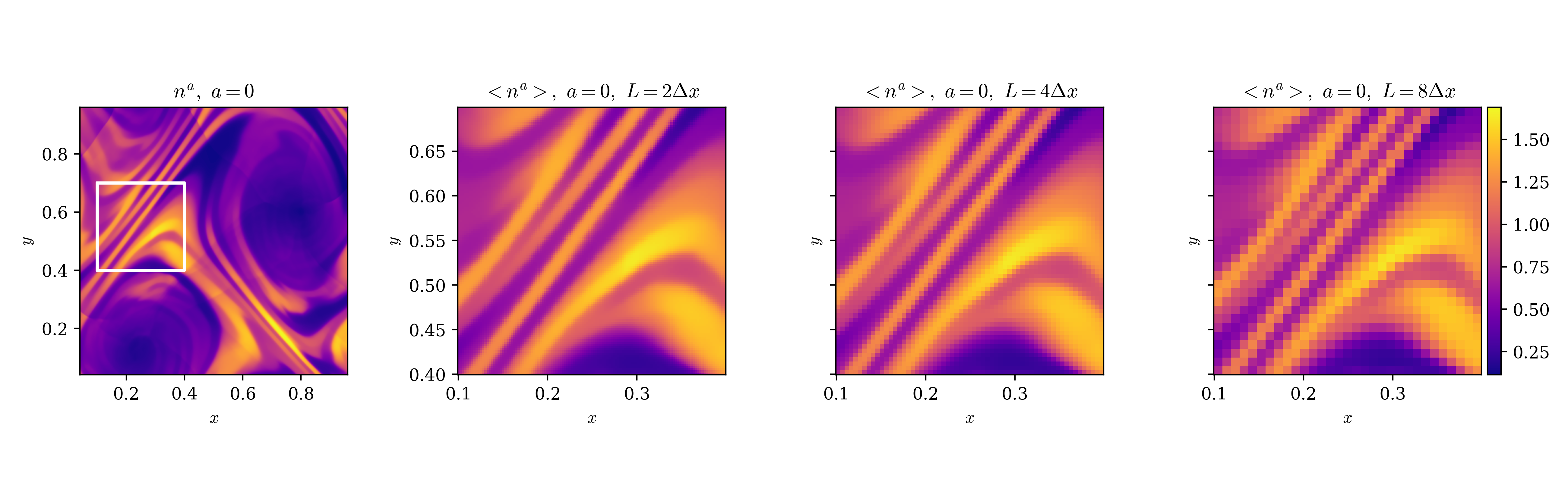}
    \caption{Comparison between fine-scale baryon current $n^a$ and filtered data using various filter-widths. The left-most panel shows a snapshot of the fine-scale data at around $t=10$. In the panels to its right we plot the filtered baryon current with filter-sizes $\qtylist{2;4;8}{\dx}$, zooming in the region $x\in(0.1,0.4),\, y\in(0.4,0.7)$.
    The corresponding patch is indicated in the left panel. Data for this figure has been produced using an observer box-length equal to the filter size and explicitly coarse-graining the meso grid.
    Coarse-graining is particularly evident in the third and fourth panel, which makes the images appear pixelated.}
    \label{fig:ComparisonCG_BC}
\end{figure}

Up until now, we have focused on individual aspects of the scheme one at a time, starting from the construction of the observers and then moving on to showing some examples of the filtered data we obtain. 
We chose to do so for two reasons: i) this is intended as a demonstration of the key points of the logic briefly presented in \cref{sec:logic}; ii) given the novelty of the strategy adopted and of the codebase written, these tests allows us to gain confidence in the validity and robustness of the analysis.
In particular, we discussed some properties of the filtering observers computed using $L = \qty{8}{\dx}$ in \cref{fig:ObsVSmicro}.
Then we considered the impact of filtering at a fixed filter-size and also varied this (while keeping the observers fixed) to show how the impact of filtering scales with the filter-size (cf. \cref{fig:FilterScalingBC}). 
In all these previous cases, the filtered data has been produced simply by interpolating data coming directly from simulations, without reducing the number of points of the meso-grid with respect to the fine-scale one. 
As such, it makes sense to conclude this section by considering the above-mentioned aspects all at once.  
In \cref{fig:ComparisonCG_BC} we show a comparison of the fine-scale and filtered baryon current varying the filter-size and, at the same time, recomputing the observers for each value of this.
Similarly, we have also increased the grid-spacing for the filtered data accordingly: if the filter-size is $\qty{2}{\dx}$ (as is the observers' box length) the number of points in the meso-grid where such data is stored is reduced by a factor of $2$ (in each direction), and so on.
We do so because our ultimate goal is to inform `low-resolution' simulations and make them (somehow) sensitive to physics happening on small scales without increasing the resolution.
In essence, we need to keep track of the information that is lost in reducing the grid-resolution.

\subsection{Stress-energy residuals and closure ingredients}\label{subsec:residuals}

\begin{figure}
    \centering
    \includegraphics[width = 1\textwidth]{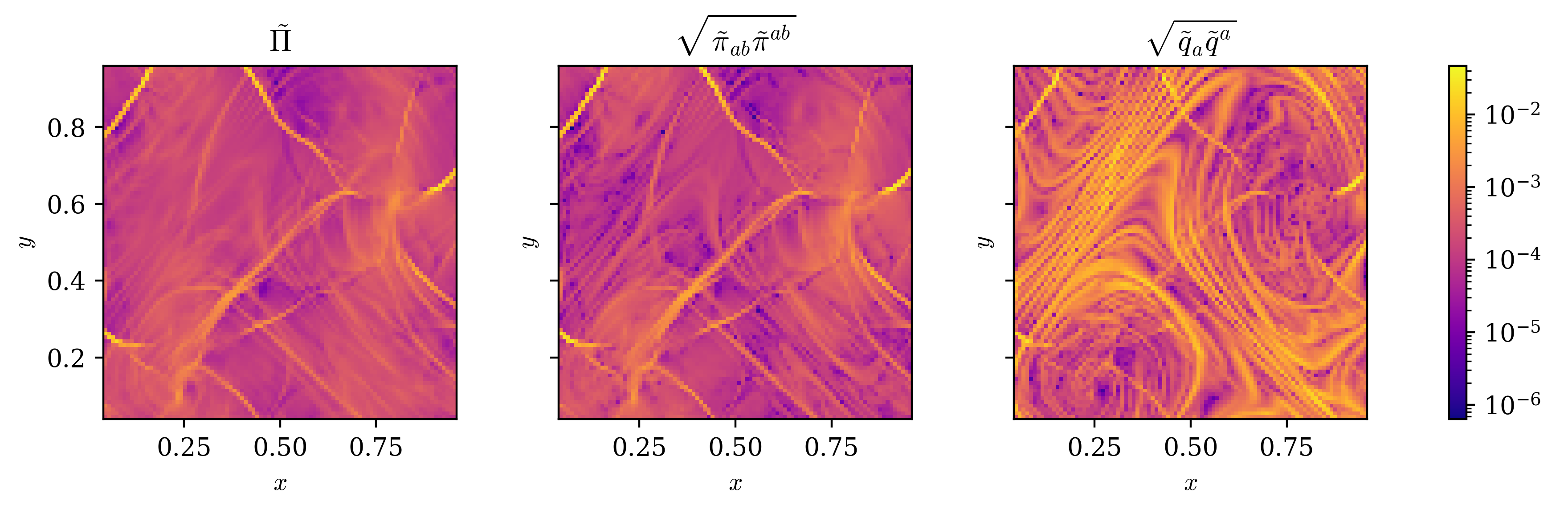}
    \caption{Plotting the magnitude of the stress-energy tensor residuals. From left to right: the isotropic residual stresses $\tilde\Pi$, the anisotropic residual stresses $\tilde\pi^{ab}$ and residual momentum flux $\tilde q^a$. Data underlying this figure is obtained from filtering a snapshot at around $t=10$ using a filter-size of $\qty{8}{\dx}$.}
    \label{fig:Residuals_cg8}
\end{figure}

Let us now turn to the modelling of the filtered/meso-scale stress-energy tensor $\la T^{ab}\ra$.
As discussed in \cref{sec:logic}, filtering introduces additional off-diagonal components in the stress-energy tensor---the residuals---which we need to model. 
We demonstrate this by first computing the Favre-observers from the filtered baryon current as
\begin{equation}
    \tn = \sqrt{- \la n^a\ra \la n_a\ra}\;, \quad \tu^a = \la n ^a \ra / \tn \longrightarrow \la n^a \ra = \tn \tu^a\;,
\end{equation}
and decomposing the stress-energy tensors with respect to these. 
In essence, we compute the residual stresses and momentum flux as 
\begin{subequations}
\begin{align}
    \tilde \Pi &= \frac{1}{3} \tperp_{ab} \la T^{ab}\ra - \la p \ra \;, \\
    \tilde \pi^{ab} &= \tperp^a_c \tperp^b_d \la T^{cd}\ra - \left(\la p \ra + \tilde \Pi\right)\tperp^{ab} \;, \\
    \tilde q^a &= - \tu_c\tperp^a_b \la T^{bc}\ra \;,
\end{align}
\end{subequations}
where we recall that $\tperp^{ab} = \eta^{ab} + \tu^a\tu^b$.
Note that the equation of state of the meso-model pressure $\la p \ra$ need not be that of the micro-model. For now we make the simplifying assumption that they match and investigate this point further in~\cref{sec:eos_check}.
An example of such residuals is given in \cref{fig:Residuals_cg8}, where once again the data is coming from a snapshot at around $t=10.0$ and the simulation data has been filtered (and coarse-grained) using a filter size of $\qty{8}{\dx}$. 
As $\tilde q^a,\,\tilde \pi^{ab}$ are tensors, in the middle and right panel of the figure we plot $\sqrt{\tilde\pi_{ab}\tilde\pi^{ab}}$ and $\sqrt{\tilde q_a\tilde q^a}$ as these give us a measure of the overall magnitude. 
We then confirm these residuals terms are generically non-negligible, as expected based on theoretical grounds \cite{celora_covariant_2021}. 

Even though we have performed the decomposition in terms of the Favre observers $\tu^a$, this is not a unique choice, certainly not from a theoretical point of view. 
We could have, for example, used the filtering observers $U^a$ for the decomposition instead, noting that this would not break covariance as the filtering observers have a clear physical meaning in our framework.
We choose to work with the Favre observers for two reasons: first, there would be additional drift terms in the filtered baryon current had we decomposed it with respect to $U^a$. These additional drift terms are essentially re-absorbed into the definition of $\tu^a$ (see \cite{celora_covariant_2021,FilteringStrategy}) and would need to be modelled otherwise. 
Second, a similarly defined Favre velocity is used in most of the Newtonian LES work on turbulence \cite{SchmidtLES}. 
In any case, given the way we construct the filtering observers, we have checked that there are no large differences between $U^a$ and $\tilde u^a$ as expected. 
Having discussed the residuals in the stress-energy tensor, let us turn to their modelling. 
To do so we first need to compute various gradients of the Favre velocity $\tu^a$ and of the temperature $\tT$ (cf. \cref{eq:EckartClosure}), noting that our closure scheme will require us to combine spatial and temporal derivatives.
This is evident if we look at the definition of the shear tensor $\tsig_{ab}$ in \cref{subeq:Favre_gradients}: the shear matrix is, by definition, spatial with respect to the Favre observer, but the observer's spatial directions are tilted with respect to the foliation. 
We stress that this is not a peculiar feature of the closure scheme we are exploring here, rather it is reasonable to expect that this will happen with \emph{any} covariant closure scheme. For example, if we want the closure scheme to account for the fact that we expect turbulent transport only if there is non-zero shear in a local Lorentz frame moving with the fluid, then this will inevitably require it to mix time and space derivatives (cf. \cite{duez_comparison_2020}). 
Having said that, we compute space and time derivatives with respect to the foliation---using standard second-order centered finite-differencing---and then from these we reconstruct, say, the velocity gradient decomposition as in \cref{subeq:Favre_gradients}.\footnote{A convenient way of doing this is to store in memory the various derivatives as components of a tensor. This way we can directly apply the algebraic decomposition as in \cref{subeq:Favre_gradients}. We also note that such decomposition makes explicit use of some algebraic constraints which may be violated due to numerical errors.
Large deviations from these constraints should be corrected for, although we checked explicitly and this appears not to be required in our case.}

Computing spatial derivatives is fairly straightforward since it requires using filtered data coming from a single snapshot. 
Time derivatives, on the other hand, require some extra work as we also need filtered data coming from different time slices: for a centred second-order finite-differencing scheme we need 3.
We also note here that the time gap between the slices play a role at this point.
Since space and time derivatives are here computed using the same differencing scheme we choose the time gap between the stored simulation snapshots to be \emph{equal} to the spatial grid spacing. 
As a first pass, we decided in this work not to interpolate the gridded data.
Therefore, having the slices separated by a time gap equal to the grid spacing implies that all derivatives are computed with the same accuracy: with the same order and the same increment.

\begin{figure}
    \centering
    \includegraphics[width = 0.96 \textwidth]{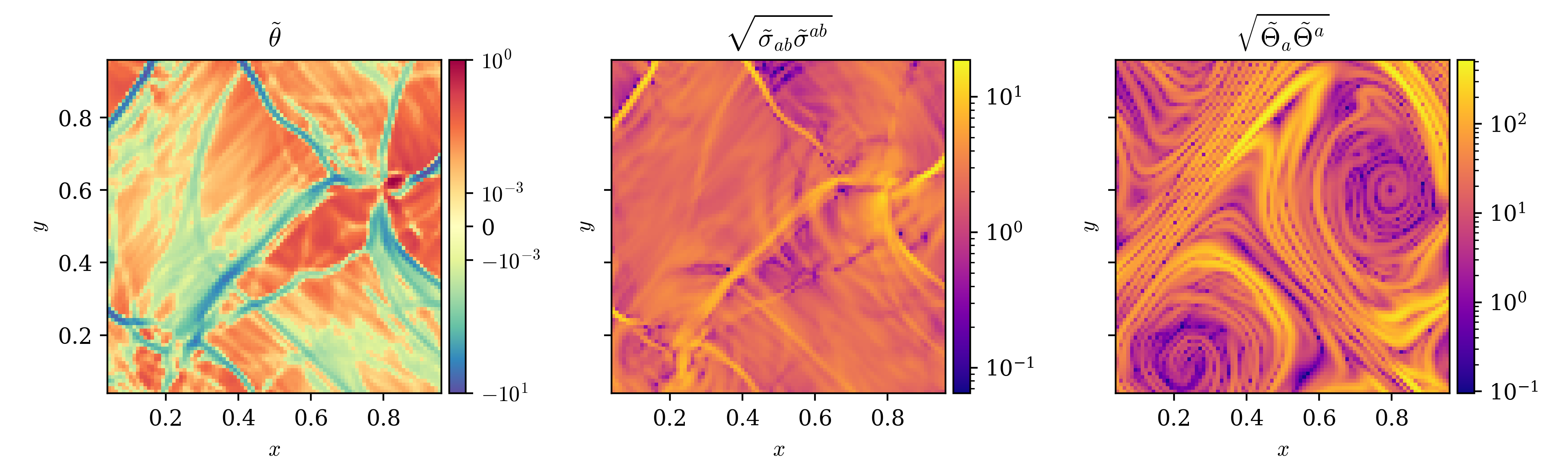}
    \caption{Results for various gradients relevant to the modelling of the stress-energy tensor residuals. From left to right we show: the expansion rate $\ttheta$, the square-root of the trace of the squared shear tensor $\sqrt{\sigma_{ab}\sigma^{ab}}$, and the temperature gradients corrected by the heat inertia $\sqrt{\tilde\Theta_a\tilde\Theta^a}$, where $\tilde\Theta^a = \tperp^{ab}(\nabla_b\tT + \tT \ta_b)$. Data for this figure was obtained from filtering a snapshot at around $t=10$ using a filter size of $\qty{8}{\dx}$.}
    \label{fig:Gradients_cg8}
\end{figure}

In \cref{fig:Gradients_cg8} we show an example of some gradients relevant to the modelling of the residuals plotted in \cref{fig:Residuals_cg8}.
In particular, in the left we show the Favre-observers' expansion rate $\ttheta$, while in the middle and right panel we plot $\sqrt{\tsig_{ab}\tsig^{ab}}$ and $\sqrt{\tilde\Theta_a\tilde\Theta^a}$, where $\Theta^a = \tperp^{ab}(\nabla_b \tT + \tT \ta_b)$.
Comparing \cref{fig:Residuals_cg8,fig:Gradients_cg8}, we immediately observe a nice degree of spatial correlation between quantities plotted in the middle and right-most panels. 
Given this, it makes sense to conclude this section by considering the simplest model for the residuals we can think of. 
Let us first consider the anisotropic residual stresses $\tilde\pi^{ab}$ and residual momentum flux $\tilde q^a$, as well as the corresponding relevant gradients $\tsig^{ab},\,\tilde\Theta^a$.
At each point on the meso-grid, we `square' these and compute their ratio as 
\begin{equation}
    r_1 =\frac{\sqrt{\tilde\pi_{ab}\tilde\pi^{ab}}}{\sqrt{\tsig_{ab}\tsig^{ab}}},\qquad \text{and } r_2=\frac{\sqrt{\tilde q_a\tilde q^a}}{\sqrt{\tilde \Theta_a\tilde\Theta^a}} \;. 
\end{equation}
We then average over the values computed at each grid point---meaning that we average the ratios to obtain $\bar r_1,\,\bar r_2$---and construct the models for the residuals as $\tilde\pi^{ab}_{mod} = \bar r_1 \tsig^{ab}$ and $\tilde q^a_{mod} = \bar r_2 \tilde\Theta^a$. 
We do the same for the isotropic residual stresses $\tilde\Pi$, noting that we construct the model for this using the absolute value of $\tilde\theta$. This is because, while $\tilde \Pi$ is positive at all gridpoints, the expansion rate is not (cf. \cref{fig:Residuals_cg8,fig:Gradients_cg8}).

\begin{figure}
    \centering
    \includegraphics[width=0.96\textwidth]{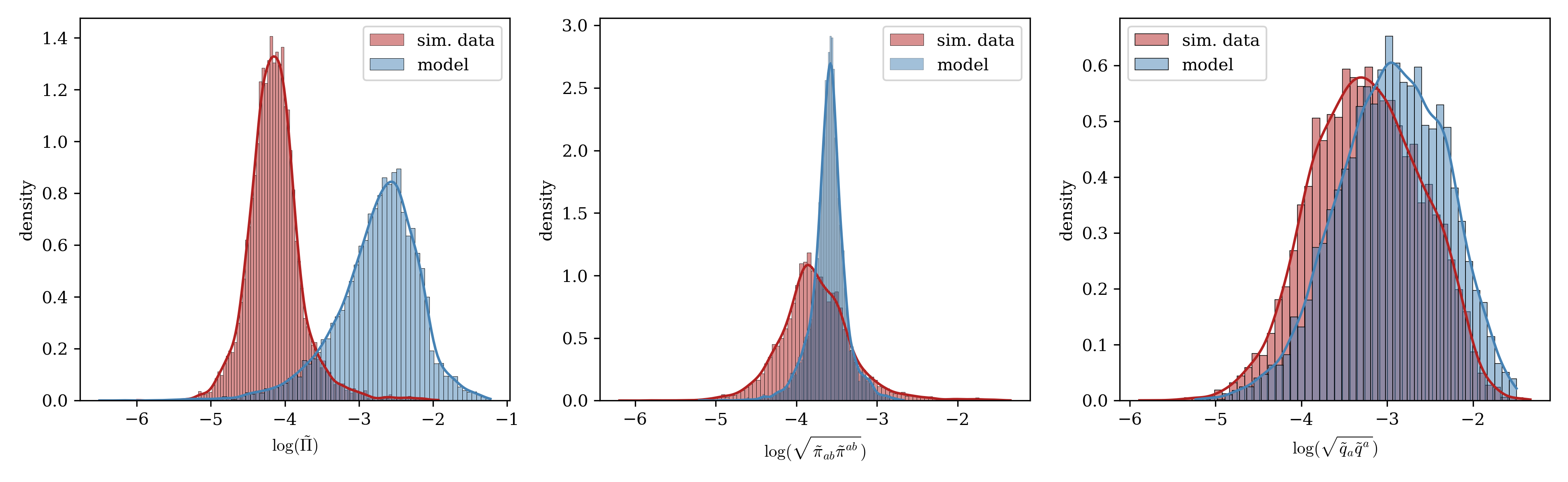}
    \caption{Comparing distributions of stress-energy tensor residuals with their values from a first-order gradient model that assumes constant coefficients. From left to right we show: the isotropic residual stresses $\tilde\Pi$, the anisotropic residual stresses $\tilde\pi^{ab}$ and the residual momentum flux $\tilde q^a$.}
    \label{fig:Const_coeff_closure}
\end{figure}

We compare the results to the residuals by plotting the corresponding distributions in \cref{fig:Const_coeff_closure}.
The first thing we note is that, while we do not expect this model to be particularly good or accurate, the model for the residual momentum flux (right-most panel of the figure) is actually better than expected: while the mean values of the distributions are slightly different, there is a nice overlap between the two. 
This is suggestive of the fact that a constant heat conductivity model might be not too far off. 
Having said that, let us turn to the isotropic residual stresses (left-most panel of the figure), which is visibly the worst case of the three.
That this was going to be the worst case can be easily explained: when comparing the left-most panels of \cref{fig:Gradients_cg8} with that of \cref{fig:Residuals_cg8}, we see that $\ttheta$ shows complex dynamics that are not perfectly represented by $\tilde \Pi$. 
In this sense, it is evident that a constant coefficient model for the isotropic residual stresses will not be particularly good. 
Also consider the anisotropic residual stresses (middle-panel of the figure).  
We observe that, while the model almost correctly captures the mean value, it is not able to reproduce either the spread nor the skewness of the distribution.
The next section will be devoted to discussing how we can improve on this  crude model. 

\section{Discriminating between models}\label{sec:DiscriminatingModels}

The aim of this section is to discuss some initial ideas on how to improve the modelling of the extracted transport coefficients and their corresponding residuals. 
We will focus on the anisotropic residual stresses, given that these are expected to play the largest role in determining the dynamics of a shearing flow like the one in the Kelvin-Helmholtz instability.
We also conducted a similar analysis for the remaining residuals but as the qualitative discussion and key messages are essentially the same,  we report on them in \cref{app:RemainingResiduals}. 

\subsection{Scaling with filter width}

\begin{figure}
    \centering
    \includegraphics[width=0.99\textwidth]{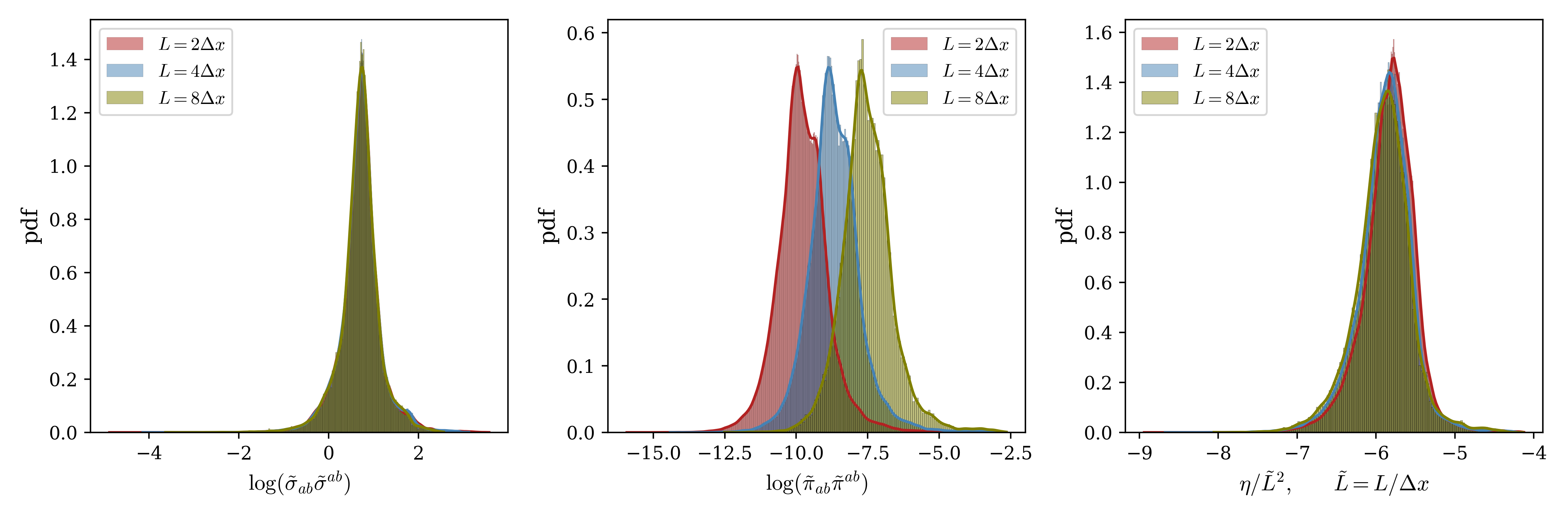}
    \caption{Comparing distributions at different filter sizes. From left to right we show: i) the second invariant of the shear tensor, ii) the `magnitude' of the anisotropic stress residuals and iii) the re-scaled, extracted shear viscosity.}
    \label{fig:eta_scaling}
\end{figure}

Perhaps the most fundamental thing we need to discuss is how the residuals and the corresponding effective dissipative coefficients change as we vary the filter size.  
In order to discuss this, we need to compare the distributions obtained with different filter sizes.
For the case of the shear viscosity, the relevant distributions to look at are those of $\tilde{\pi}_{ab}\tilde{\pi}^{ab}$ and $\tsig_{ab}\tsig^{ab}$.
These are shown in \cref{fig:eta_scaling} for the three filter sizes considered in this work, $L=\qtylist{2;4;8}{\dx}$. 
Quite notably, while the distributions of $\tilde{\pi}_{ab}\tilde{\pi}^{ab}$ change by a constant shift in log space, those of $\tsig_{ab}\tsig^{ab}$ do not.
In particular, the shift between the $L=\qty{2}{\dx}$ and $L= \qty{4}{\dx}$ appears to be the same as that between the $L= \qty{4}{\dx}$ and $L=\qty{8}{\dx}$ distributions---the shift being roughly $1.2$ in log-space.
As for the shear viscosity coefficient, we compute this (as before) at each grid point as $\eta = \sqrt{\tilde\pi_{ab}\tilde\pi^{ab}}/\sqrt{\tsig_{ab}\tsig^{ab}}$.
Given that the distributions for $\tsig_{ab}\tsig^{ab}$ do not change with the filter size---hence corroborating the idea that this quantity is measuring an intrinsic property of the flow---we then expect the $\eta$-distributions to scale as\footnote{This is because $1.2 \approx \log_{10}(16)$.} $(L/ \Delta x)^2$. 
The right panel of \cref{fig:eta_scaling} confirms this  guess as we observe the scaled distributions to almost perfectly overlap.
We stress that this scaling is consistent with naïve expectations based on Taylor-expansion type reasoning, although cannot be taken for granted given the way we implement filtering here. 
It at least represents a strong test of the robustness of the framework that we are exploring here.

\subsection{Linear regression in log-space}

Having discussed the residuals' scaling with the filter size, let us turn to their modelling. 
Because the extracted values of $\eta$ range over $3$ orders of magnitude, we will pursue a linear regression model in log-space.

As a simple attempt, we can propose modelling $\eta$ in terms of only the filtered temperature and baryon density, such that $\eta \equiv \eta(\tT,\tn)$. This parametrization of a transport property in terms of thermodynamic variables is a common one motivated by viscosity's microphysical origins.
It turns out, however, that this gives very poor results: while we do not show this here, we compared the distributions of model predictions against data---both for the shear viscosity distribution as well as for the corresponding residual distribution---and observed very poor matching. 
This can be corroborated with a simple correlation plot amongst the variables $\{\eta,\tT,\tn\}$, which highlights the weak correlations between these quantities. 
This may be due to the fact that the filtered temperature $\tT$ and density $\tn$ span a range of only one order of magnitude.
This is, however, common to most simulations of Kelvin-Helmoltz instabilities we are aware of as the simulation set-up we used is quite standard---e.g. uniform pressure in the initial data.
We can try to explain this result by the fact that the Kelvin-Helmholtz instability---used here to sustain the development of turbulence---is inherently a shear-instability whose dynamics are largely controlled by the relative velocity of the two fluids. 

Given this, we expand the list of possible explanatory variables used in our regression, including in particular various scalar quantities involving gradients of the Favre-velocity $\tu^a$.  
This choice is partly motivated by the observation that some of these quantities are much better correlated with the extracted shear viscosity and partly by the fact that we might expect these quantities to actually play a role on theoretical grounds. 
The full list of possible explanatory variables considered here is
\begin{equation*}
    \left\{\tT,\, \tn,\,\tsig_{ab}\tsig^{ab},\,\text{det}(\tsig),\,\tvort_{ab}\tvort^{ab},\,\tsig_{ab}\tsig^{ab} -\tvort_{ab}\tvort^{ab},\, \tsig_{ab}\tsig^{ab}/\tvort_{ab}\tvort^{ab} \right\}\;.
\end{equation*}
In addition to the thermodynamic quantities, we include the invariants of the shear tensor, $\text{det}(\tsig),\,\tsig_{ab}\tsig^{ab}$, that of the vorticity tensor, $\tvort_{ab}\tvort^{ab}$, as well as the difference and the quotient between the second invariants of these. 
We choose to add the last two quantities in the list since the former can be used to identify a vortex in a turbulent flow \cite{Qcriterion}, whilst the latter is a measure of the relative intensity of the shear and vorticity and has recently been used in a discussion of the local-energy cascade in Newtonian hydrodynamic turbulence \cite{LocalCascade}. 
Moreover, we can make sense of the need to consider such explanatory variables in the following way: the extracted effective transport coefficient capture the energy flowing to or from the
scales we have explicitly filtered out, which we expect to happen as non-linearities kick in. 
Because we extracted these coefficients using formulae inspired from the modelling of viscous fluids and laminar flows, we can expect the intrinsic non-linearity of turbulence to manifest itself by having transport coefficients depending on velocity gradients as well.  

Given this list of possible explanatory variables we construct the best model in the following way.
We consider all possible sets and subsets of the $7$ quantities listed above and for each we run a linear regression routine. 
Splitting the data into training and validation---using a ratio of 80:20---we identify the best model by contrasting the distributions of extracted data with the model predictions. 
We use the first Wasserstein distance, also known as Earth mover's distance, to quantify the quality of a given model\footnote{We note that similar results can be obtained using Pearson's correlation coefficient as a quality factor, although we found it less capable of discriminating between models.}: given two distributions $X$ and $Y$ the Wasserstein distance between them can be computed as 
\begin{equation}
    W_1(X, Y) = \sum_i ||X_{(i)} - Y_{(i)}|| \;.
\end{equation}
In essence, once we construct the regression models, we look for the one that minimizes the distance between the data and the model predictions in a distributional sense. 

\subsection{Model interpretation and viability}

\begin{figure}
    \centering
    \includegraphics[width=0.96\textwidth]{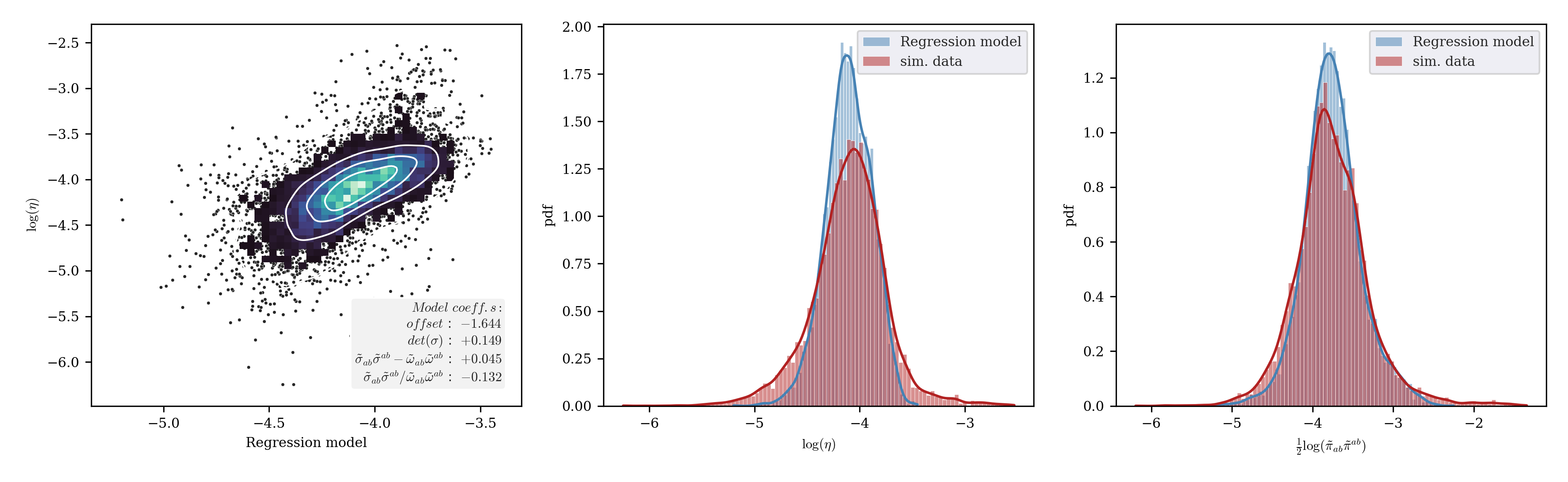}
    \caption{Best regression model in log space for $\eta$. From left to right we show: i) scatter-plot of the extracted data vs. model predictions ii) probability distributions of extracted $\eta$ vs its modelling iii) probability distributions of the extracted residual vs its modelling.}
    \label{fig:eta_best}
\end{figure}

In \cref{fig:eta_best} we summarize the results of this analysis. 
The first thing we note is that the best model we obtain does not involve thermodynamical quantities at all: the model involves only three of the quantities listed above, namely $\text{det}(\tsig),\,\tsig_{ab}\tsig^{ab} -\tvort_{ab}\tvort^{ab},\, \tsig_{ab}\tsig^{ab}/\tvort_{ab}\tvort^{ab}$. 
This is reported in the left panel of \cref{fig:eta_best}, where we show a scatter plot between the model and the data, while the annotation box reports the regression coefficients of the model. 
The middle panel of the figure instead shows a comparison between the distributions of the model and the data, where we observe that this model does not fully capture the longer tails of the data's distribution. 
Finally, we show in the right panel of the figure a comparison between the actual residual under consideration and the model's predictions. A strong matching between the two is seen.
For the sake of clarity, let us comment on how we build the model's predictions for the residuals. We consider the validation set and for each sample point we build the prediction for $\eta$ given the pointwise values of the model's regressors. Then we construct the residual predictions as $\tilde \pi^{ab}_{model} =\eta_{model}\,\tsig^{ab}$. 
Ultimately, the results shown in the right panel are the ones we truly care about. We see that the two distributions largely overlap. However, this simple model cannot quite explain the skew in the distribution data, nor its slightly longer tails. 
Nonetheless, the improvement with respect to the constant coefficient case above is significant and can be appreciated by comparing this panel with the middle one in \cref{fig:Const_coeff_closure}.

The data underlying \cref{{fig:eta_best}} has been filtered using a filter-size of $L=\qty{8}{\Delta x}$. 
It is also important, however, to check how the best model constructed in this way changes as we vary the filter size---both in the set of explanatory variables used and in the specific values of the coefficients. 
We have then performed the same process on data filtered with sizes $L= \qtylist{2;4}{\Delta x}$ and found that the best set of explanatory variables is unchanged across the different cases and the regression coefficients take almost identical values. 
The only feature that changes is the offset, which increases by $\approx 0.6$ as we double the filter size. 
This is consistent with the $(L/\Delta x)^2$ scaling shown in \cref{fig:eta_scaling} and provides a useful check on the robustness of the proposed modelling.  

\change{It is also important to add a word of caution on the well-posedness of this scheme as this is not well understood. 
Certainly the available results in the literature do not cover the case where the transport coefficients are non-linear functions of the velocity gradients. 
At the same time, we may expect that if the correction terms are small enough, then in practice we would be close to modelling ideal hydrodynamics, so that well-posedness may be inherited. 
In any case, the model explored here is intended as a first stab to the problem, and more advanced ones that are free from such potential problems can be introduced and calibrated following the same logic. 
}

We conclude this section by commenting on an additional requirement for a model such as this to be viable. 
In fact, we need to make sure that the model we construct is regular in the laminar limit. 
This is also linked to the issue of asymptotic preservation, and we return to it in \cite{FilteringStrategy}.
Given the best model above, this would mean that the value for $\eta$ should not diverge in the limit where $\tsig\to 0$.
At a first glance, this might appear problematic since the regression coefficient for $\tsig_{ab}\tsig^{ab}/\tvort_{ab}\tvort^{ab}$ is negative. 
Nonetheless, we recall that $\text{det}(\tsig) \sim \tsig^3$, while $\tsig_{ab}\tsig^{ab}\sim \tsig^2$. As such, given that the coefficient in front of the $\text{det}(\tsig)$ regressor is positive and larger than that of $\tsig_{ab}\tsig^{ab}$, we assert that the model constructed is actually regular in the laminar limit. 

\section{A closer look at the implications of the Equation of State residual}
\label{sec:eos_check}

So far in this work we have repeatedly mentioned, and then by-passed, the fact that filtering may impact on the thermodynamics, as well.
In particular, we mentioned the fact that EoS residuals may introduce a disconnect between the microphysics and the EoS input in a numerical simulation.
While we return and expand on such issue in \cite{FilteringStrategy}, the aim of this section is to fill in this gap by taking a closer look at the equation of state residuals. 
This will also allow us to take the first step towards understanding the implications this may have if not accounted for.

To begin with, let us look back at \cref{eq:FilteredTab_decomposition} and note that there are two residual contributions entering in the total trace of the filtered stress-energy tensor. 
We see a first one coming from the trace of $\tilde s^{ab}$, which we model in \cref{app:RemainingResiduals}, plus a second arising from the equation of state non-linearities. 
In fact, following from the discussion in  \cite{celora_covariant_2021}, we can write the filtered pressure as 
\begin{equation}
    \la p\ra = - \teps + \tilde\mu \tilde n + \tilde T \tilde s + M \;.
\end{equation}
where $M$ is the EoS residual, $\teps,\,\tn$ are the energy and number density (measured by the Favre observer $\tu^a$), while $\tilde T,\, \tilde \mu$ are defined with respect to a meso-scale EoS of the form $\tilde p = \tilde p (\tn,\teps)$.
For example, if we choose to work with a `$\Gamma$-law' EoS as in \cref{eq:GammaLawEoS}, the meso temperature is obtained as $\tT = \tilde p /\tilde n$.
As non-linearities in the Gibbs relation can be interpreted as `entropy-like' contributions associated with the fluctuations, there is some freedom in the choice of the filtered thermodynamic potential (e.g. the filtered entropy $\tilde s$).
This choice will affect the extracted chemical potential $\tmu$ and temperature $\tT$ and, in turn, the EoS residual $M$. 
The most natural choice would be to work with the same functional form for the equation of state, but this is a-priori a non-unique choice \cite{FilteringStrategy}. 

\begin{figure}[!htb]
    \begin{minipage}{0.26\textwidth}
        \centering
            \includegraphics[scale = .38]{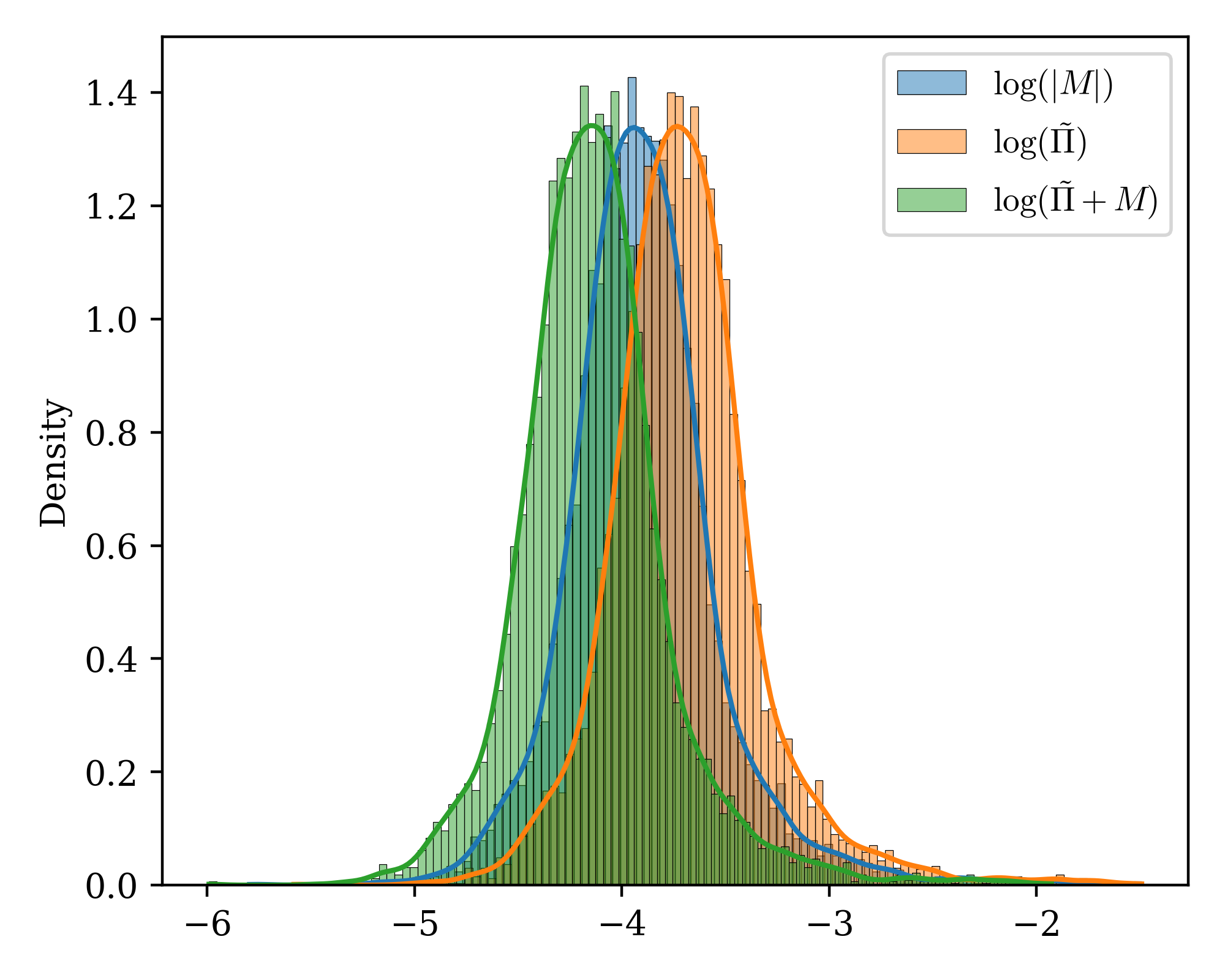}
    \end{minipage}
    \begin{minipage}{.73\textwidth}
        \centering
            \includegraphics[scale=0.41]{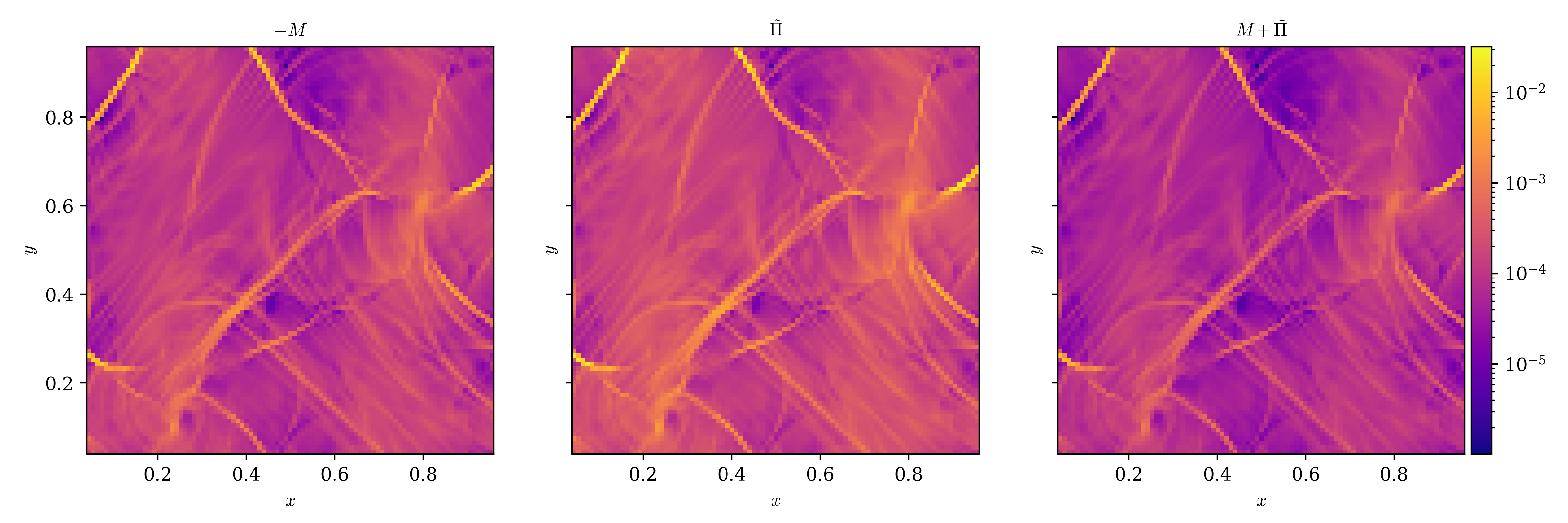}
    \end{minipage}
\caption{Visualizing the different residuals contributing to the trace of the stress-energy tensor. In the left-most panel we show distributions corresponding to the EoS residual, the isotropic stresses residual and the sum of the two. In the panels to its right we plot the same quantities individually. 
Data used in this figure has been filtered with filter-size $L=\qty{8}{\dx}$ using snapshots around $t=10$.}
    \label{fig:EOSres_combined}
\end{figure}

First of all, we consider the individual residual contributions to the trace of the filtered stress-energy tensor.
These are plotted in \cref{fig:EOSres_combined} for a snapshot at around $t=10$ and filtering with size $L=\qty{8}{\dx}$. 
For this figure we chose to work with the same EoS as used for the micro-model, that is a `$\Gamma$-law' EoS.
We then observe that for this choice the EoS residual $M$ is always negative, so we plot in \cref{fig:EOSres_combined} the absolute value. 
In contrast, the $\tilde \Pi$ residual is always positive
as is their sum, so that we have a net positive increase in the pressure term. 
This can also be appreciated looking at the relative distributions in the left-most panel of \cref{fig:EOSres_combined}. 

\begin{figure}
    \centering
    \includegraphics[width = 0.96\textwidth]{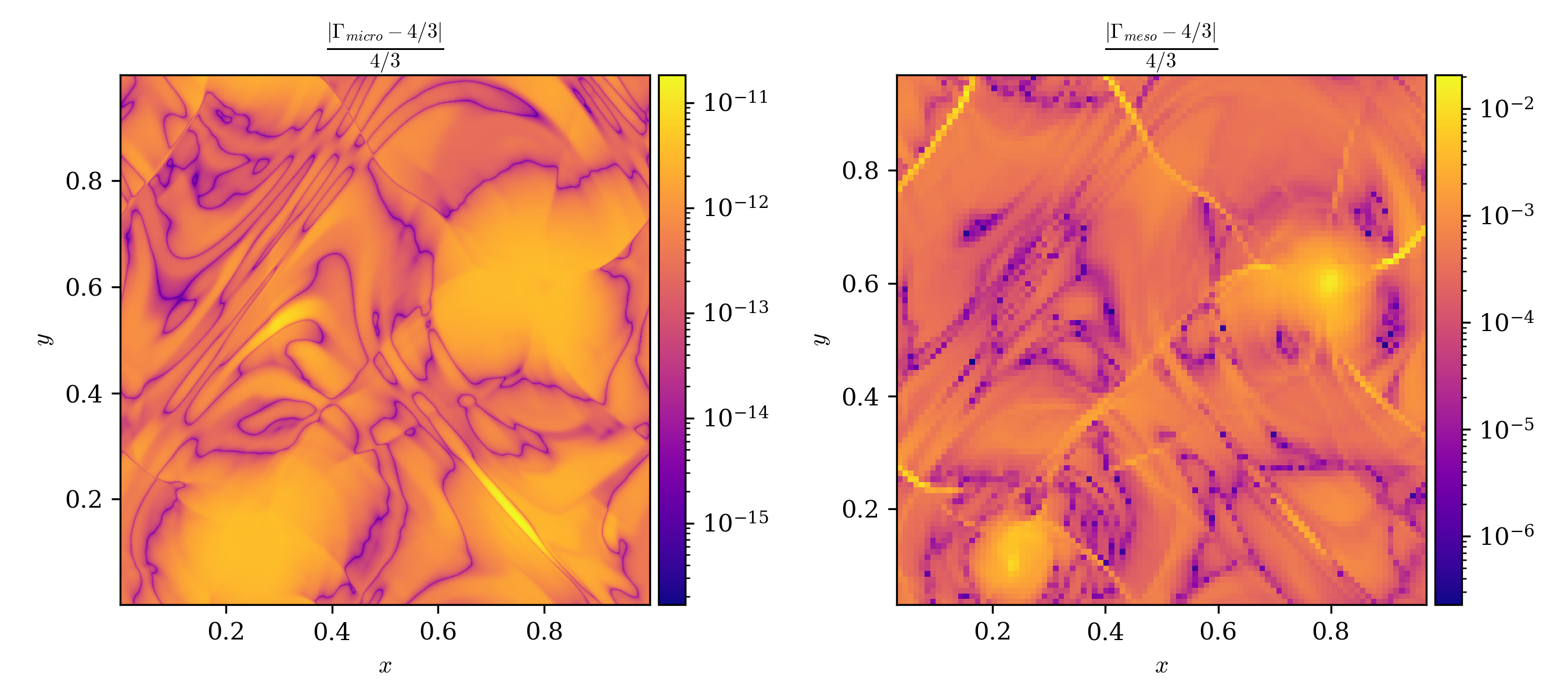}
    \caption{Comparing the first adiabatic index $\Gamma_1$, computed locally using \cref{eq:CalculabelGamma1}, to the expected value of $4/3$. The left panel corresponds to the fine-scale data, with a largest relative difference on the order of $10^{-11}$. The right panel corresponds to the filtered data, explicitly ignoring the EoS residual. For this case we observe differences up to percent level. The data underlying this figure has been filtered with filter size $L=\qty{8}{\dx}$ using snapshots around $t=10$.}
    \label{fig:DeviationsGammaLocal}
\end{figure}

Having seen that the EoS residuals $M$ are generically non-negligible---their magnitude is of the same order as the others---it is natural to ask the `null-hypothesis' question: what is the impact of neglecting this? 
This question is relevant in a broader sense as the effect of filtering will inevitably enter any numerical simulation due to the implicit filtering associated with the numerical discretization. 
As such we want to investigate whether these residuals can have a `measurable' impact on the EoS parameters. 
To do so we extract locally the effective adiabatic index 
\begin{equation}
    \Gamma_1 = \left(\frac{\partial \log p}{\partial \log n}\right)_{x_s}
\end{equation}
where $x_s$ is the specific entropy, but first we need to rewrite this in terms of variables we have access to. 
Choosing to work with the energy and number densities $n,\,\veps$ we arrive at
\begin{equation}\label{eq:CalculabelGamma1}
    \Gamma_1 = \frac{n}{p} \left[\left(\frac{\partial p}{\partial n}\right)_\veps + \left(\frac{\partial p}{\partial \veps}\right)_n\left(\frac{\partial \veps}{\partial n}\right)_{x_s}\right]  = \frac{n}{p} \left[\left(\frac{\partial p}{\partial n}\right)_\veps + \frac{p + \veps}{n}\left(\frac{\partial p}{\partial \veps}\right)_n\right]  \, .
\end{equation}
As a quick sanity-check one can verify that for a $\Gamma$-law equation of state this expression gives $\Gamma_1=\Gamma$ as expected.
Then we consider the (corresponding) values for $p,\veps,n$ explored in the METHOD simulation and construct a smooth bi-variate cubic spline approximation for $p(n,\veps)$.
With this we can then take derivatives and evaluate \cref{eq:CalculabelGamma1} at each grid-point. 
We do so for the micro-model data---that is the simulation output of METHOD---and compare it to the value used in the numerical code, $\Gamma=4/3$.
In  the left panel of \cref{fig:DeviationsGammaLocal} we plot the relative difference between the two and observe that this is at maximum of the order of $10^{-11}$.

As for the meso-model, we do the same but assuming the null hypothesis, that is pretending that $M$ is not there. 
We then follow the same logic as above but now in terms of the filtered pressure $\la p \ra$, energy density and number density $\teps,\tn$.
Assuming that $M$ has no impact at all is equivalent to say that $\la p\ra$ is related to $\teps,\,\tn$ by the same equation of state as at the micro-level.
We test this by extracting $\Gamma_1$ locally using \cref{eq:CalculabelGamma1}---although replacing now $p \to \la p\ra,\, \veps \to \teps,\, n\to \tn$---and comparing it to the null-hypothesis value of $4/3$. 
The relative difference between the two is plotted in the right panel of \cref{fig:DeviationsGammaLocal}. 
In sharp contrast with the values obtained for the unfiltered data, we observe differences up to the \emph{percent level}. 

Let us  try to spell out what possible consequences this may have by considering a binary neutron star merger simulation with finest resolution of $\O(\qty{10}{\metre})$.
The results shown suggest that unmodelled turbulent dynamics happening on scales of $\O(\qty{1}{\metre})$ could induce up to percent level errors in the EoS parameters.
While we would still need to quantify how much of a bias this introduces in EoS inferences and parameter estimation, it is clear this is an issue we need to explore  in more detail. 
This is even more true if we consider that we expect turbulent dynamics to be induced at much smaller scales than $\O(\qty{1}{\metre})$ in an actual merger, see \cite{DavidIanReview}.
Having turbulent dynamics at smaller scales than $\O(\qty{1}{\metre})$ would mean that we can consider the $\O(\qty{10}{\metre})$-simulation as effectively introducing a filter with size larger than $\approx \qty{10}{\dx}$. 
As the residuals---including the EoS one, $M$---scale with the filter-size, the potential systematic biases this may introduce could be even larger than reported here.

\section{Conclusions and Outlook}

Binary neutron star mergers are extremely rich and dynamical events whose quantitative modelling requires using numerical relativity simulations, particularly for the latest stages of the inspiral, the actual merger and the post-merger phase. 
The expected sensitivities of future third generation detectors, as well as the fact that these will also be able to detect the post-merger gravitational wave signal, justifies the ongoing efforts to improve the realism of said simulations.
A key issue in this respect is the need to resolve (at least in principle) the full range of scales involved in the turbulent flow that develops in the merger remnant---e.g. due to the Kelvin-Helmholtz instability \cite{kiuchi_efficient_2015,priceRosswog2006}. 

Because this is (at the very least) not practical, recent years have witnessed various efforts to extend the large-eddy simulations' strategy from Newtonian physics to relativistic (magneto-)hydrodynamics \cite{radice_calibrated,carrasco_gradient_2020,vigano_general_2020}.
While these efforts have already delivered impressive results \cite{aguilera-miret_turbulent_2020,aguilera-miret_universality_2021,palenzuela_turbulent_2022,aguilera-miret_role_2023,izquierdo_large_2024}, there are a number of non-trivial aspects involved that are not yet fully understood.
Certainly one of these is the fact that all practical implementations so far break covariance, both in the way the filtering operation is performed and in the specific closure scheme used \cite{DavidIanReview}.

In this respect, in \cite{celora_covariant_2021} we have discussed some of these issues and put forward a coherent theoretical framework for performing large-eddy simulations while retaining compatibility with the covariance principle of general relativity (a higher-level discussion of the reasons for tackling such an issue in this way is provided in \cite{FilteringStrategy}). The obvious advantage of this approach is that subgrid schemes tuned on high-resolution turbulence results in special relativity can be directly lifted to general curved spacetimes. The disadvantage is that the subgrid model will need recalculation for every change in the micro- and meso-model parameters, including any change in the choice of equation of state. Considering these implementation aspects further will be left for future work.

We have taken a practical step forward by presenting the very first implementation of a fully covariant filtering scheme in relativity. 
Being the first of its kind, this work is primarily intended as a demonstration that the logic discussed in \cite{celora_covariant_2021} is practically viable. 
We presented simulations of Kelvin-Helmholtz driven turbulence and used these to discuss the key features of the scheme.
In particular, we dynamically identify  a suitable observer that moves with the bulk of the flow and perform the filtering in the spatial directions identified by this observer.
We demonstrate that, as expected from theoretical grounds, filtering impacts on the stress-energy tensor by introducing effective dissipative terms and that these scale with the size of the filter kernel as naïvely expected. 
As an illustration of the viability of the scheme, we have then provided a first `a-priori' calibration of the effective transport coefficients that can be extracted directly from simulation data using a simple eddy-viscosity-type model. 
Finally, we have looked at the impact that filtering may have on the thermodynamics, finding indications that, if not properly modelled, this may introduce up to percent level differences in the EoS parameters. These results demonstrate the kind of questions we may try to answer by working within this framework.

The purpose of this work is also to introduce and present the codebase we have built\footnotemark{}
In this respect, it is important to stress that the results presented here are not expected to be of direct use for real large-scale simulations of compact-object mergers. 
This is mainly because the simulations used here have two spatial dimensions, whereas it is well known that three-dimensional turbulence is phenomenologically quite different from its two-dimensional counterpart. 
Nonetheless, we stress that the key routines have been written in such a way to work in $3+1$ dimensions as well. 
Similarly, the scheme has been structured in a pipeline-like fashion to easily allow for future improvements and extensions. 
For example, future work will range from extensions to account for electromagnetic effects---the relevant theory aspects to extend the current implementation to magnetized systems are discussed in \cite{FilteringStrategy}---to a systematic exploration of various closure schemes and `a-posteriori' tests of these. \change{In particular, `a-posteriori' tests are needed in order to fully assess the actual performance and degree of universality of any hydrodynamic turbulence model.}
To this end, we note that recent work by \citet{Hatton2024DEIFY} provides a computationally efficient formulation of dissipative hydrodynamics that can be used to implement a sub-grid model of the type we derive here.

\acknowledgments
TC is an ICE Fellow and is supported through the Spanish program Unidad 
de Excelencia Maria de Maeztu CEX2020-001058-M.
NA and IH gratefully acknowledge support from Science and Technology Facility Council (STFC) via grant numbers ST/R00045X/1 and ST/V000551/1.
The authors acknowledge the use of the IRIDIS High Performance Computing Facility and associated support services at the University of Southampton.
TC also acknowledges the use of the HIDRA cluster and related support, at the Institute of Space Sciences (ICE-CSIC). 
TC would like to thank D. Viganò for various useful discussions during the completion of this work.
We acknowledge the use of the following open-source software: \textsc{hdf5} \cite{HDFGroup_2024}, \textsc{MATPLOTLIB} \cite{hunter_matplotlib_2007}, \textsc{NumPy} \cite{harris2020array}, \textsc{SciPy} \cite{virtanen2020scipy}, \textsc{seaborn} \cite{Waskom2021Seaborn}, \textsc{sk-learn} \cite{pedregosa2011scikit}.

The first release of code used for generating these results is publicly 
available on GitHub.\footnotemark[\value{footnote}]
\footnotetext{The first release of the code used in this work is publicly available at \href{https://github.com/Lagrangian-filtering/Lagrangian-filtering}{https://github.com/Lagrangian-filtering/Lagrangian-filtering}\label{ft:github}}

\begin{appendix}

\section{Isotropic stresses and momentum flux residuals modelling}\label{app:RemainingResiduals}

\begin{figure}
    \centering
    \includegraphics[width=0.96\textwidth]{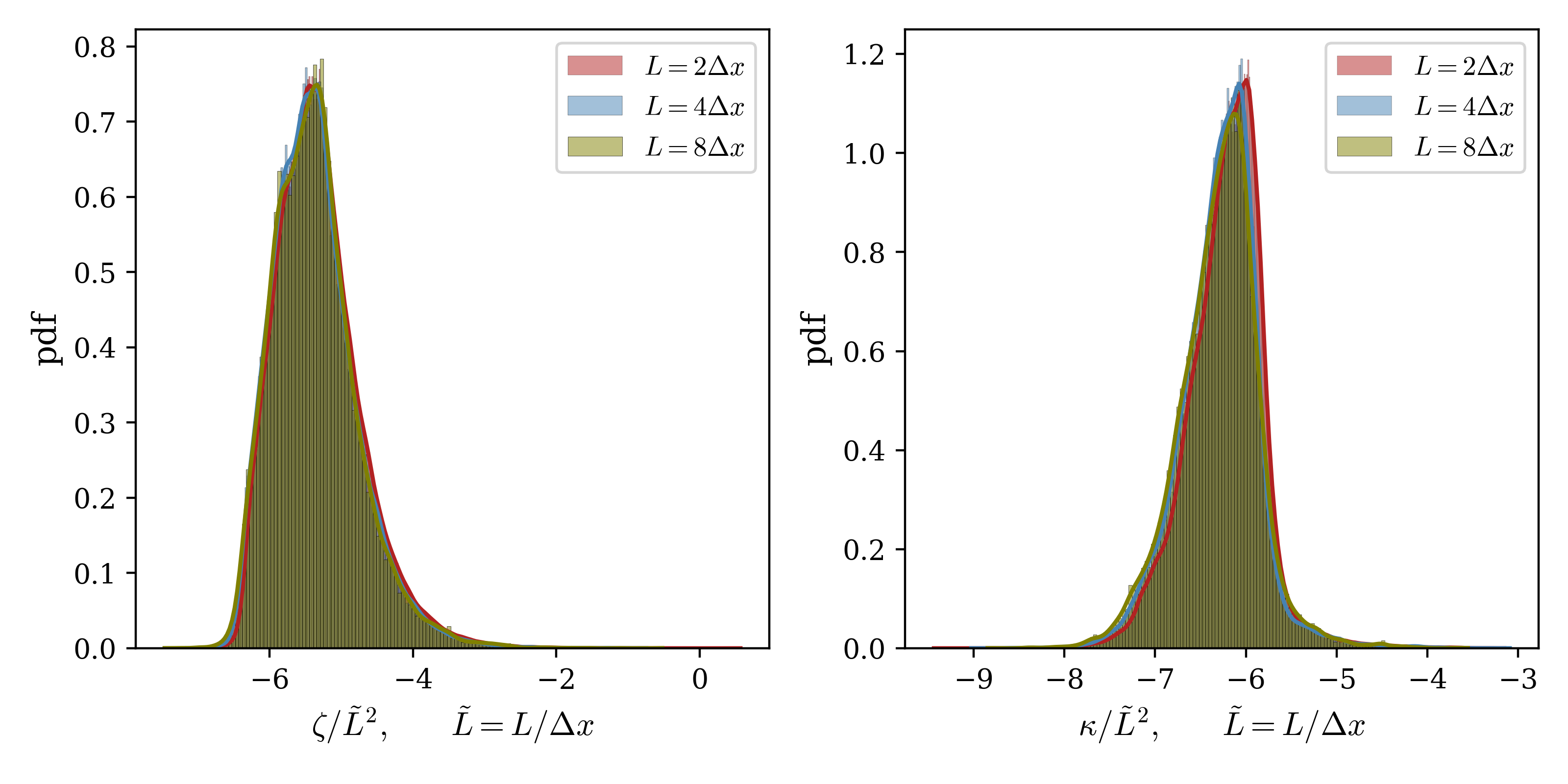}
    \caption{Comparing distributions of $\zeta$ (left) and $\kappa$ (right) at  different filter sizes. The almost perfect overlap of the re-scaled distributions demonstrate that the coefficients (and the residuals) follow the same scaling as discussed for $\eta$.}
    \label{fig:remaining_residuals_scaling}
\end{figure}

In this Appendix we report on the modelling of the isotropic stresses and momentum flux residuals, following the same logic as discussed in the main body of this work for the anistropic stresses.
First, we observed the distributions for the remaining residuals to scale with the filter-size in the same fashion as for the anisotropic stresses: doubling the filter-size corresponds to a rigid shift of the distributions to larger values (the shift being $\approx 1.2$ in log-space as above).
Second, the distributions corresponding to the relevant gradients---the expansion rate $\tilde \theta$ and the temperature gradients $\tilde \Theta_a \tilde \Theta^a$---do not change as we varied the filter-size.
As in the case of the effective shear viscosity then, we expect the distributions of $\zeta$ and $\kappa$ to scale with $(L \unit{\dx})^2$. 
We confirm this by plotting in \cref{fig:remaining_residuals_scaling} the corresponding re-scaled distributions as we find these are practically indistinguishable.

\begin{figure}
    \centering
    \includegraphics[width=0.96\textwidth]{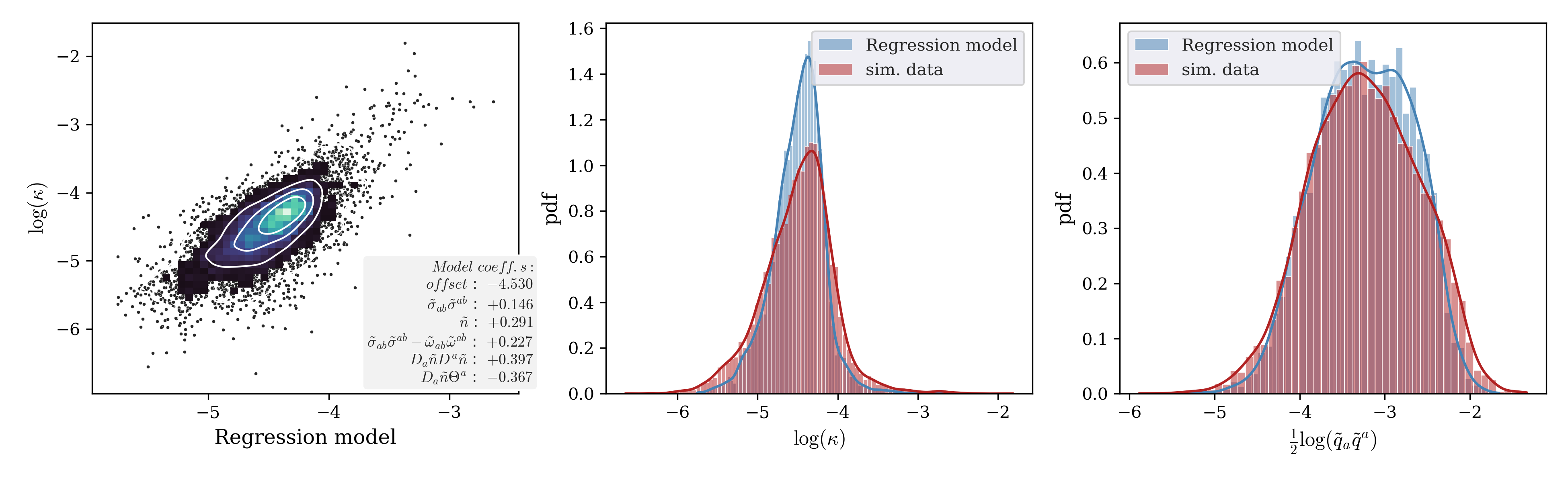}
    \caption{Equivalent to \cref{fig:eta_best} but focusing on the heat conductivity. From left to right we show: i) scatter-plot of the extracted data vs. model predictions ii) probability distributions of extracted $\kappa$ vs its modelling iii) probability distributions of the extracted residual vs its modelling.}
    \label{fig:kappa_best}
\end{figure}

\begin{figure}
    \centering
    \includegraphics[width=0.96\textwidth]{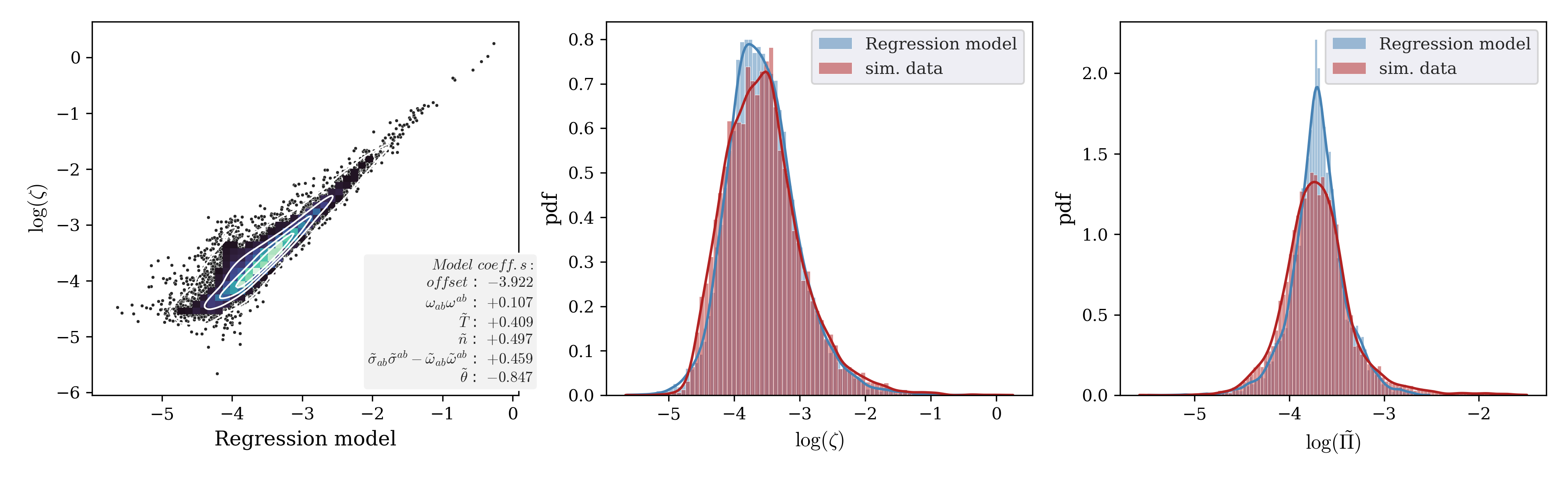}
    \caption{Same as \cref{fig:eta_best} but focusing on the bulk viscosity. From left to right we show: i) scatter-plot of the extracted data vs. model predictions ii) probability distributions of extracted $\zeta$ vs its modelling iii) probability distributions of the extracted residual vs its modelling.}
    \label{fig:zeta_best}
\end{figure}

When it comes to the modelling, the only modification with respect to the shear viscosity case is that now we include additional quantities in the list of possible residuals: the new ones considered here are 
\begin{equation*}
    \{\ttheta, \ta_a\ta^a,\,\tilde{\Theta}_a\tilde{\Theta^a},\,(D_a\tn)(D^a\tn),\,(D_a\tn) \tilde\Theta^a, \dot\tn, \dot\tT\}\,. 
\end{equation*}
These have been added as we found spatial gradients of the number density and the temperature to be effective at modelling the effective heat conductivity (cf. \cref{fig:kappa_best}).
As for the bulk viscosity, we unsurprisingly found that adding the expansion rate to the list of regressors significantly improved the model's performance (cf. \cref{fig:zeta_best}).
We also note that the best-fit exponent obtained for the expansion is close to $-1$.
In practice then, the dependence of the isotropic residuals $\tilde\Pi$ on the expansion rate almost cancels out.
This is consistent with the poor correlation observed between $\tilde\Pi$ and $\ttheta$ and casts some doubts on the validity of interpreting the isotropic residuals as an effective bulk viscous pressure. 
Noting that this is not necessarily problematic, we also stress that in order to reach a firm conclusion would require using more data as well as more refined statistical models (and/or closure schemes). 
Given the scope of this work we leave this for the future.

\section{Extracting \texorpdfstring{$\eta$}{the shear viscosity coefficient} componentwise}

In the main body of this work, we extracted the magnitude of the shear viscosity coefficient by `squaring' the relation in \cref{eq:EckartClosure}.
In a sense, it would be natural to ask what would happen had we extracted the coefficient differently. 
For example, one could think of extracting $\eta$ using the same relation as above but in a component-wise fashion: this would mean extracting 5 more independent distributions that eventually need to be combined in some way.  

However, it is important to stress at this point that (in our view) this appears to be somewhat problematic, particularly because of the effort we made to set up the observers and perform the filtering in a covariant fashion. 
Phrasing this differently, we prefer the `squaring' option as it allows us to stay `as covariant as possible', while extracting $\eta$ component-wise is at odds with one of the main arguments for performing the filtering in the way we do it in this work. 
Nonetheless, and mainly because we find this to be instructive, let us expand here on the results we get if we extract $\eta$ from each component of \cref{eq:EckartClosure} independently.

\begin{figure}
    \centering
    \includegraphics[width = 0.7\textwidth]{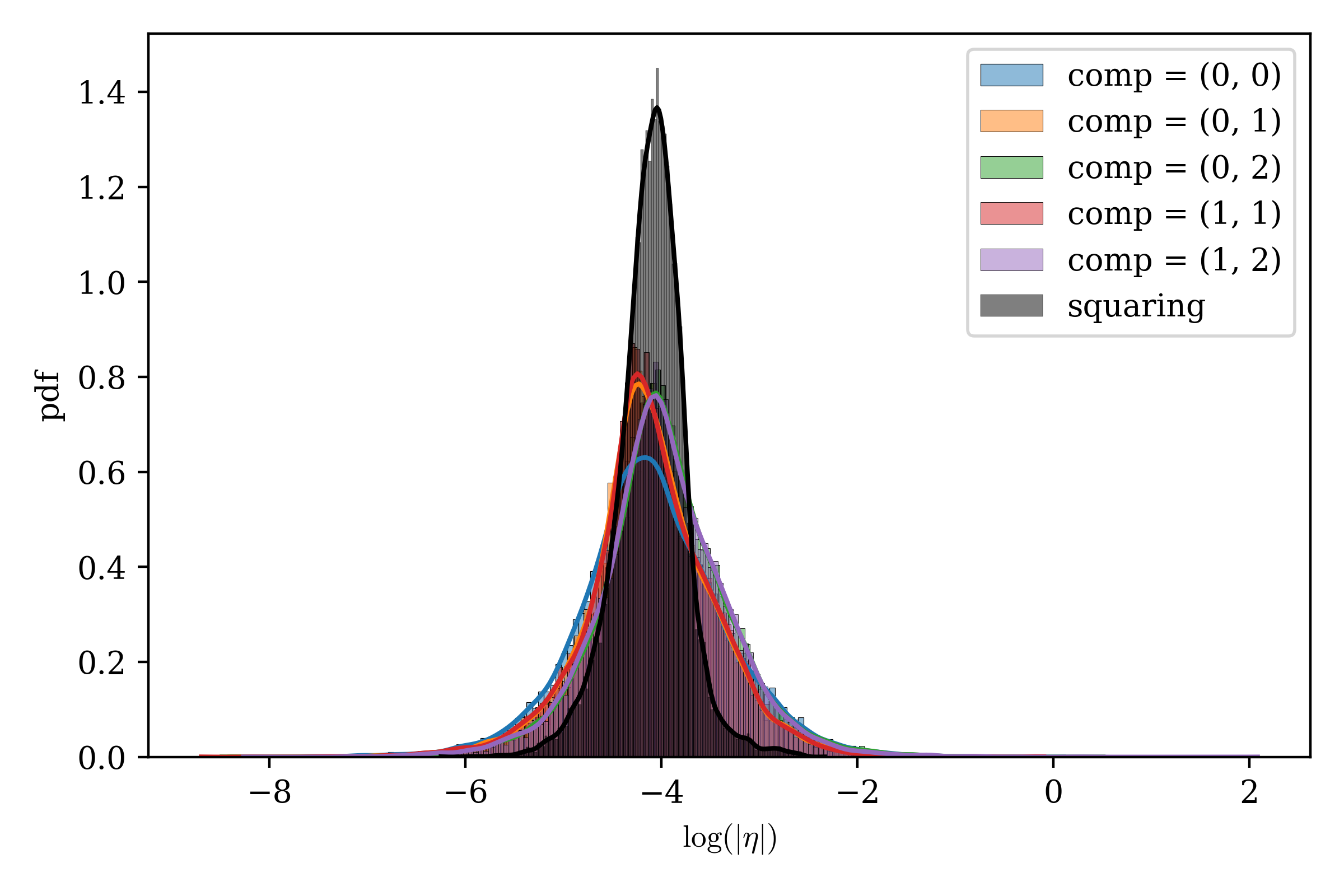}
    \caption{Extracting $\eta$ component-wise: comparing the magnitude distributions obtained in a component-wise fashion (one per independent component) vs that via squaring. Data underlying this figure has been filtered with filter-size $L=\qty{8}{\dx}$ using snapshots around $t=10$.}
    \label{fig:Eta_componentwise_comparison}
\end{figure}

In \cref{fig:Eta_componentwise_comparison} we compare the distributions for the absolute values of $\eta$ obtained via `squaring' vs component-wise. 
The first thing we note is that the various component-wise distributions are qualitatively similar to each other. 
We take this as an indication of the fact that while we expect numerical errors to have a larger impact on the component-wise distributions---for example, when a particular component of the shear tensor is small but, say, its second invariant is not---these are somehow distributed isotropically in a statistical/distributional sense. 
Second, when comparing the component-wise distributions against that obtained via squaring, we note that they all have the same mean but the latter has smaller variance.
Stated differently, the component-wise distributions have longer tails with respect to the one obtained via squaring---which may be due to the impact of the above-mentioned numerical errors. 
While finding significantly different distributions would not be inherently problematic, the fact that we do not is reassuring. 
In fact, we may even suggest that the values obtained via squaring are reasonably consistent with those obtained in component-wise fashion while being free from the coordinate dependence of the latter.

Having said that, we feel the urge to comment on the sign to be assigned to $\eta$. As discussed in \cite{celora_covariant_2021}, the sign of the effective transport coefficients cannot be constrained or fixed using thermodynamics arguments based on the second law of thermodynamics. 
In fact, we may even say that a positive shear viscosity corresponds to a positive energy cascade \emph{locally}, where the energy is transferred to smaller scales, with the opposite being the case for negative values. 
It is also clear that in extracting the coefficient via `squaring' we are only considering its magnitude, while the sign is not a priori fixed. 
We would then need a procedure/logic to assign the sign that is similarly covariant. 
One reasonable way of doing this is the following: we can diagonalize the anisotropic stress residuals $\tilde\pi^{ab}$ and consider the direction associated with the maximum eigenvector (in absolute value). 
We could then transform the shear tensor to the basis given by the eigenvectors of $\tilde\pi^{ab}$, check whether $\tsig^{ab}$ is aligned or anti-aligned to $\tilde\pi^{ab}$ in this particular direction and assign a value to $\eta$ accordingly. 
Arguably, this would be a geometric way of fixing the sign. 

\begin{figure}
    \centering
    \includegraphics[width = 0.95\textwidth]{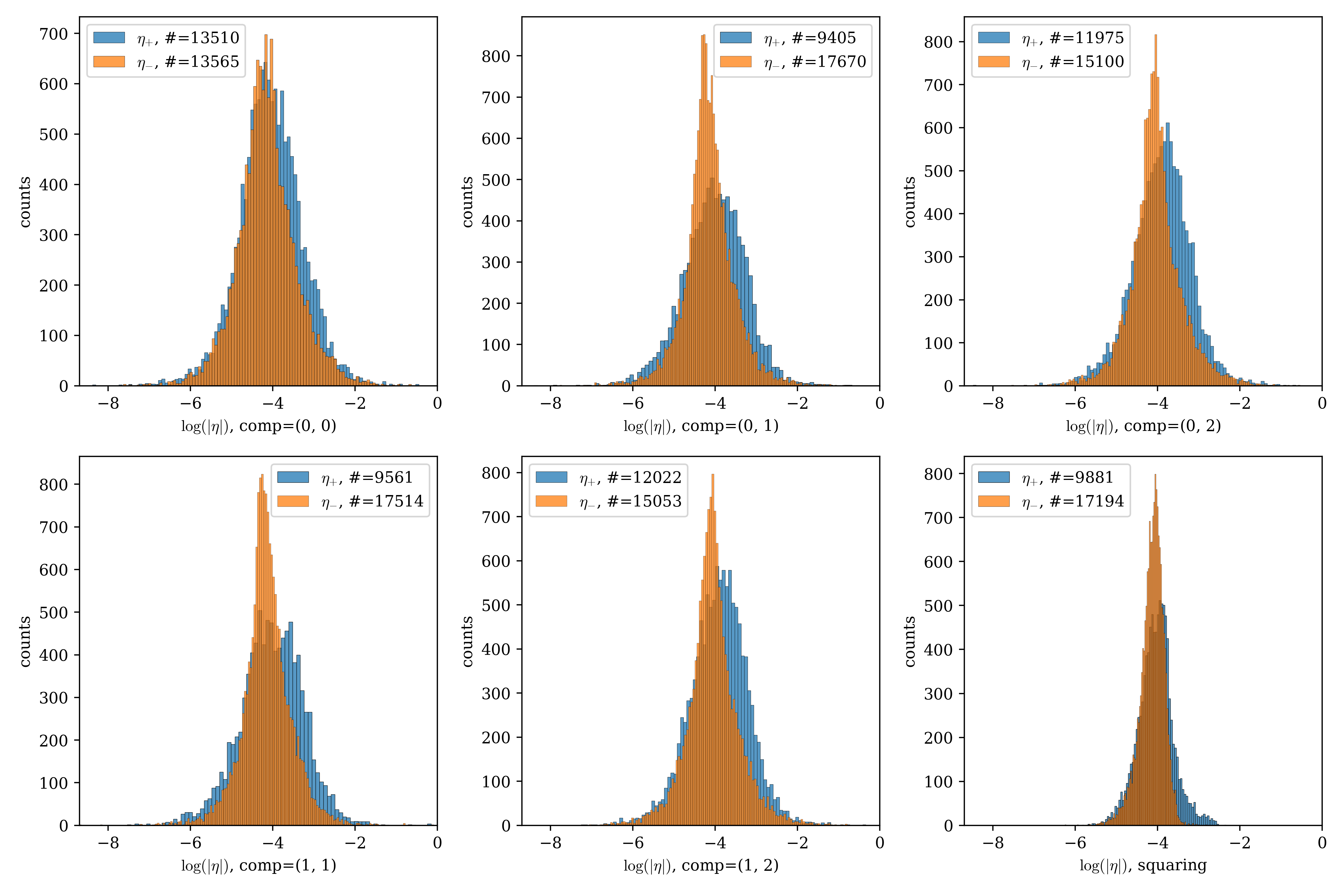}
    \caption{Comparing the histograms for positive and negative values of $\eta$, component-wise and via squaring. 
    For each panel we report the total counts of positive vs negative values, so to be able to appreciate whether there are more positive than negative values or vice versa.
    Data underlying this figure has been filtered with filter size $L=\qty{8}{\dx}$ and snapshots around $t=10$.}
    \label{fig:Eta_componentwise_pos_neg}
\end{figure}

In \cref{fig:Eta_componentwise_pos_neg} we show the distributions of positive and negative values obtained this way and compare them to what we obtain if we follow the component-wise procedure. 
While the distributions differ for different components, we observe that those corresponding the $(0,1),\,(1,1)$ and $(0,2),\,(1,2)$ components are reasonably similar. 
We also observe generically that distributions of positive values have a larger spread. 
In the figure we chose to plot the histograms and also report in the legend the positive and negative counts: we can observe then a larger number of negative values in all cases with the exception of the $(0,0)$ distributions. 
The bottom-right panel of the figure shows instead the distributions obtained via squaring and the logic described above for fixing the sign. 
Also in this case we observe a distribution of positive values with a larger variance and an overall larger number of negative values. 
In a loose sense then, this can be considered as a confirmation of the correctness of the logic used for fixing the sign, given that it also gives a larger amount of negative values that are generically smaller in magnitude. 

We conclude this appendix by commenting on why we focused on modelling the magnitude of the shear viscosity only in \cref{sec:DiscriminatingModels}. 
In fact, it is not quite clear how seriously we should take the negative values extracted:  some works in the literature simply discard these arguing they represent conservative spatial fluxes at the filter scale rather than interscale interactions (see e.g. \cite{vela2022subgrid}), even though the recent analysis in  \cite{LocalCascade} appears to suggest we may want to take them seriously. 
We also stress that the simulations used here are 2+1 dimensional, while for real applications we will need to perform the analysis on data coming from 3+1 simulations.
Besides, the phenomenology of 2-dimensional turbulence is quite different from the 3-dimensional case. In particular, we expect real 3-dimensional isotropic and homogeneous turbulence to have a net positive energy cascade.
Assuming this would correspond to an overall predominance of positive over negative values---as reported in \cite{LocalCascade}---and given the proof-of-concept nature of this work, we leave the exploration of these aspects for the future. 

\end{appendix}

\bibliography{biblio.bib}

\end{document}